\documentclass[np2,edit2,twoside,twocolumn,showpacs,superscriptaddress]{revtex4-2}

\usepackage[utf8]{inputenc}
\usepackage{comment}
\usepackage{amsmath}
\usepackage{amssymb}
\usepackage{enumerate}
\usepackage{bm, ulem}
\newcommand{\bt}[1]{\mathbf{#1}}
\usepackage[pdftex, pdftitle={Article}, pdfauthor={Author},colorlinks,citecolor=blue,linkcolor=blue,urlcolor=blue]{hyperref}
\usepackage{cleveref}
\usepackage{xcolor}
\usepackage{graphicx}
\usepackage[export]{adjustbox}
\begin{document}

%\preprint{APS/123-QED}

\setcounter{page}{1}
\title[]{Inducing $\mathbb{Z}_2$ Topology in Twisted Nodal Superconductors}

\author{Kevin P. \surname{Lucht}}
\affiliation{%
 Department of Physics and Astronomy, Center for Materials Theory,
Rutgers University, Piscataway, NJ 08854, USA
}%
\author{Pavel A. \surname{Volkov}}
\affiliation{%
Department of Physics, Harvard University, Cambridge, Massachusetts 02138, USA
}%
\affiliation{%
Department of Physics, University of Connecticut, Storrs, Connecticut 06269, USA
}%
\author{J. H. \surname{Pixley}}
\affiliation{%
 Department of Physics and Astronomy, Center for Materials Theory,
Rutgers University, Piscataway, NJ 08854, USA
}%
\affiliation{
Center for Computational Quantum Physics, Flatiron Institute, 162 5th Avenue, New York, NY 10010
}%
\date{\today}

\begin{abstract}

Twisted nodal superconductors have been shown to exhibit chiral topological superconductivity under broken time-reversal symmetry. Here we show how a time-reversal preserving topological superconductivity can be induced in nodal triplet superconductor multilayers. For a bilayer system, the application of a Josephson spin current in triplet superconductors induces a non-zero spin Chern number per node in momentum space. However, we show that stabilizing a nontrivial global $\mathbb{Z}_2$ invariant requires an odd number of layers.  As a specific example we consider trilayers with three forms of twist: chiral, alternating and single-layer. For chiral and single-layer case, we find that a gap opening in the dispersion leads to a non-trivial $\mathbb{Z}_2$ topological invariant. For single layer twists, we show how this invariant is non-trivial when extended to an arbitrary odd number of layers.
\end{abstract}

\maketitle
%\tableofcontents

%%%%%%%%%%%%
%%%%%%%%%%%%
\section{Introduction}
%%%%%%%%%%%%
%%%%%%%%%%%%

Twisted nodal superconductors are a promising platform to realize topological superconductivity through the spontaneous or the induced formation of a chiral order parameter \cite{Can2021,Volkov2022,Margalit2022}. Of  potential superconducting candidate materials, $\textrm{Bi}_2\textrm{Sr}_2\textrm{CaCu}_2\textrm{O}_{8+\delta}$ (BSCCO) has received considerable attention due to its %van der Waals bond between each layer 
highly two-dimensional (2D) structure,
remaining superconducting  even in the limit of a  single layer (comprising two CuO$_2$ planes) with a similar transition temperature as in  its bulk counterpart~\cite{Yu2019}. Although twisted bilayers have not been realized, twisted flakes have been achieved leading to the observation of time-reversal symmetry breaking (TRSB) interfacial superconductivity
~\cite{Volkov2021,Tummuru2022_2,martini2023twisted,zhao2023time}.
As experimental techniques advance, various forms of stacked and twisted  cuprates from one to several layers with the necessary nodal structure are a plausible achievement in the near future. Beyond twisted cuprates, other layered systems have been predicted theoretically to exhibit topological characteristics. Some of these predictions include chiral superconductivity using layered triplet superconductors~ \cite{tummuru2020chiral,Volkov2022_2,liu2023},
heterobilayers with higher order topology~\cite{Li2023}, or
effective triplet superconductors formed via singlet superconductors~\cite{Lin2023}.

The focus of the bulk of the proposals above has been using TRSB to form $\mathbb Z$ topological superconductors  (classified as class C or D in the Atland-Zirnbauer (AZ) classification~\cite{Altland1997}). However, in 2D, time-reversal invariant (TRI) topological superconductors can  exist with a $\mathbb Z_2$ index for class DIII~\cite{Schnyder2008,Qi2009,Ryu2010,Qi2010}.
This class of superconductors is mainly reserved to triplet superconductors as a singlet order parameter preserves the SU$(2)$ spin symmetry, although various proximity effect arrangements~\cite{Deng2012,Zhang2013,Qin2022} or external fields~\cite{Zhang2021} can yield a TRI topological phase without the need for a triplet superconductor. 

In analogy to previous work on chiral topological superconductivity~\cite{Volkov2022}, we will explore here the possibility to induce $\mathbb{Z}_2$ superconductivity in twisted nodal superconductors by external perturbations that preserve time-reversal symmetry. In particular, spin current is one such perturbation that will be a major focus of the following manuscript.
%Although twisted nodal superconductors is a contemporary topic, 
Spin currents have been extensively studied for their long lifetime in ferromagnetic and primarily singlet superconducting interfaces~\cite{Brydon2008,Alidoust2010,Hikino2013,Costa2020,Dai2022}. Experimentally, spin-dependent supercurrents are realizable in such interfaces with surprisingly large coherence lengths, such as CrO$_2$~\cite{Keizer2006,Anwar2010,Singh2015,Han2020Review}. Recently, other magnetic interfaces with superconductors
have also been studied such as trilayer interfaces with $\textrm{La}_{1-x}\textrm{Sr}_{x}\textrm{MnO}_3$~\cite{Sanchez-Manzano2022,Kumawat2023}, $\textrm{Fe}_3\textrm{Ge}\textrm{Te}_2$~\cite{Hu2023}, and anti-ferromagnetic interfaces of Mn$_3$Ge \cite{Jeon2018}. For triplet superconductors, theoretical studies have shown that misalignments of the $d$-vector of the order parameter~\cite{Asano2006,Rashedi2007} or with the magnetic moments of a magnetic separation layer~\cite{Brydon2009} lead to spin current, and in particular non-dissipative (Josephson) ones \cite{Asano2006}.

\begin{figure*}
    \centering
    \includegraphics[width=.95\textwidth]{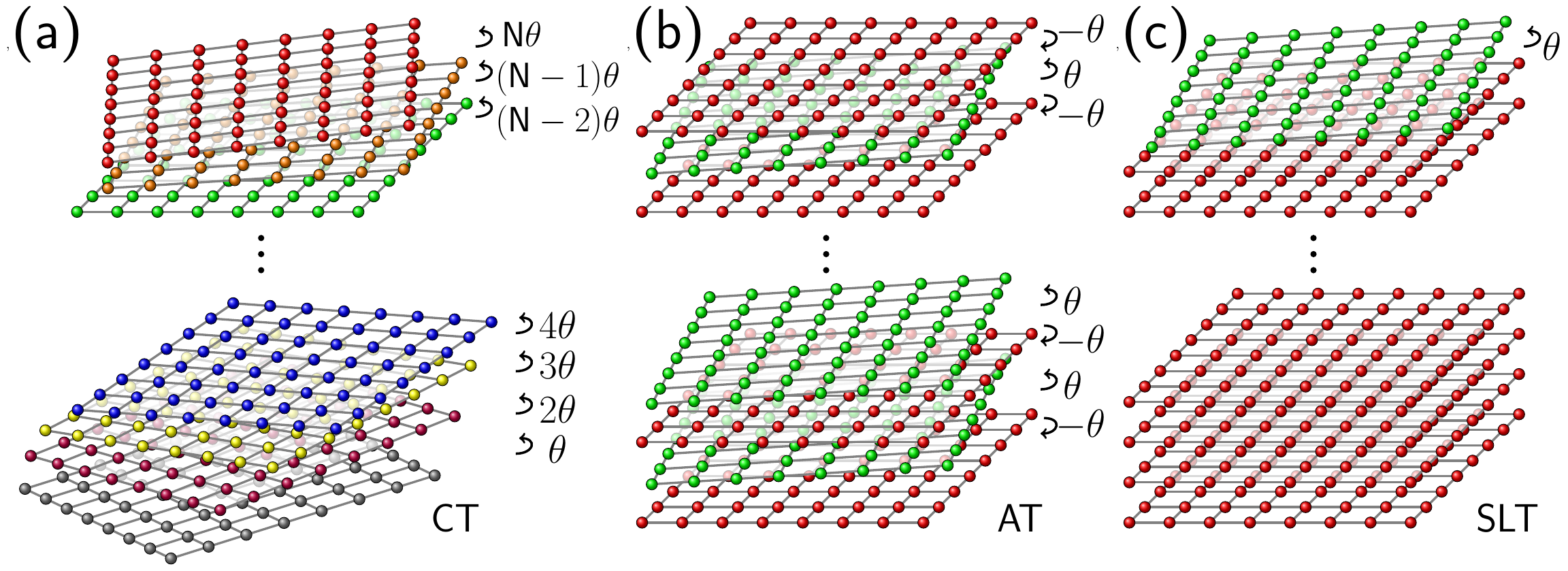}
    \caption{Twisting arrangements for $N$ layered systems. At the top of each twisting arrangement is the $N=3$ layered case which are extended below to arbitrary $N$. The arrow indicates the direction a given layer is rotated by an angle $\theta$. (a) displays chiral twists (CTs) where each layer $n$ is rotated by angle $n\theta$ to form continually rotated layers with a fixed orientation. (b) displays an alternating twists (ATs) where odd layers are rotated by angle $\theta$ and even layers by $-\theta$. (c) displays a single layer twist (SLT) arrangement where layer $N$ is rotated at an angle $\theta$ while the remaining $N-1$ layers are not twisted.
    }
    \label{fig:twisting_arrangements}
\end{figure*}

In this work, we demonstrate a strategy to create time-reversal preserving $\mathbb{Z}_2$ topological superconductors in twisted multilayers of nodal superconductors. In particular, we show that a Josephson spin current
 opens a topological gap in the spectrum which is characterized by a spin Chern number. The total spin Chern number of the system forms a  $\mathbb{Z}_2$ invariant and we show that it can be non-trival only for  systems with an odd number of layers. As an explicit example, we use a three layer system (that we extend to $N$-layers) to show how the topological character depends on the twisting arrangement, which is no longer unique beyond a pair of layers. We focus on three common twisting arrangements displayed in Fig.~\ref{fig:twisting_arrangements}: where the twist angle between nearest neighboring layers is constant and with the same sign dubbed chiral twists (CT) or with an alternating sign called alternating twists (AT) and where only the top  layer is twisted with the remaining layers untwisted that we have dubbed single layer twists (SLTs) and is the main focus of the present work. While the case of AT is topologically trivial~\cite{Tummuru2022}, for  
 the CT and SLT systems with an odd number of layers, a topological gap is formed with a non-trivial  topological index.
 Focusing on the SLTs for a three layer system, we explicitly show in a small twist angle model how to compute the $\mathbb Z_2$ index, and then extend this to an arbitrary number of layers to show its stability with an odd number of layers.

 The proceeding manuscript is organized as follows: in Sec.~\ref{sec:sym}, we examine the topological classification of a general superconducting Hamiltonian and perturbations of a $N$ layered system. We attribute one of these perturbations to a spin Josephson effect which is explored in Sec.~\ref{sec:JSC} using a two layered system which leads to a trivial topological invariant. Sec.~\ref{sec:3layer_twists} extends this spin Josephson effect to a three layered system to show that in combination with their arrangement has a nontrivial topological invariant. We conclude with Sec.~\ref{sec:fin} and comment on the topological invariant in a general $N$ layered system and its connection to 2.5-dimensional (2.5-D) systems~\cite{lucht2023}.

%%%%%%%%%%%%
%%%%%%%%%%%%
\section{Symmetry Classification of Continuum Model}
%%%%%%%%%%%%
%%%%%%%%%%%%
\label{sec:sym}

To identify in general what twisted superconductors are topological, we utilize the Cartan symmetry classifications focusing on the superconducting Altland-Zirnbauer (AZ) symmetry classes. To realize $\mathbb{Z}_2$ topology we have to extend the single layer classification of 2D superconductors  to an $N$ layered system, and then identify perturbations which allow for the formation of a TRI $\mathbb Z_2$ topological superconductor for layered systems. 
In this extension, we will first ignore the influence of twist angle, and then introduce contributions from arbitrary small twist angles between layers.

\begin{figure}
\centering
%\begin{minipage}[t]{.5\textwidth}
    \includegraphics[width=.9\columnwidth]{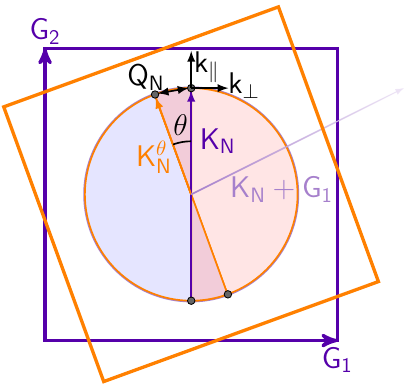}
%\end{minipage}
%
%         \includegraphics[width=.4\textwidth]{./figures/2layer_dispersion_JSC_circular.pdf}
%    \caption{Spin Josephson current in a bilayer system. Shown is a cartoon depiction a two nodal $p$ wave superconductors rotated by an angle $\theta$. The contours represent the Fermi surface with the color its sign. The location of the Dirac nodes are denoted by orange dots. By applying equal and opposite interlayer spin currents ($I^{\sigma\sigma}_{JSC}$ where $\sigma$ refers to the spin channel), a spin Josephson current arises. Consequently, a simultaneous rotation in Nambu and spin space occurs which generates a mass gap and helical edge modes represented by the indicated cartoon inset. These helical edge modes can be composed as a spin Chern number and subsequently a $\mathbb{Z}_2$ topological index.  }
    \caption{An over-view of the Brillouin zone of a twisted interface with circular Fermi surfaces. The 
    interior color represents the sign of the superconducting order parameter with line of zeros displayed as orange and purple.  The intersection of the line of gap zeros with the Fermi surface (black circles) depict the Dirac nodes.  The nonrotated layer in purple has reciprocal lattice vectors $\bt{G}_1$ and $\bt{G}_2$ and node positioned at $\bt{K}_N$. From $\bt{K}_N$, a local coordinate system is constructed. When the top layer in orange is rotated by an angle $\theta$, its node positioned at $\bt{K}^\theta_N$ introduces a displacement vector $\bt{Q}_N$ along $k_\perp$. Higher order tunneling represented by the fading vector $\bt{K}_N+\bt{G}_1$ are neglected. }
    \label{fig:coords}
\end{figure}

The second quantized Hamiltonian for  a $N$ layered system in the continuum can be written as
\begin{eqnarray}
    & \mathcal{H} = \sum_{\bt{k}}\Psi^\dagger_{\bt{k}} H_N(\bt{k}) \Psi_{\bt{k}},
    \label{eq:H_general}
\end{eqnarray}
where we use Balian-Werthammer (B-W) spinors $\Psi^\dagger_{\bt{k}} = (\Psi^\dagger_{\bt{k},1},\Psi^\dagger_{\bt{k},2},\dots,\Psi^\dagger_{\bt{k},N})$ and $\Psi^\dagger_{\bt{k},l}  = ( c^\dagger_{\bt{k},\uparrow,l}, c^\dagger_{\bt{k},\downarrow,l}, c_{-\bt{k},\uparrow,l}, c_{-\bt{k},\downarrow,l} )$  for each layer $l=1,2,\dots, N$. This Hamiltonian describes a single valley where the momenta $\bt{k}$ are local to a Bogoliubov-de Gennes (BdG) Dirac node positioned at $\bt{K}_N$ such that $\bt{k} = \bt{K} - \bt{K}_N$ where $\bt{K}$ is an arbitrary momenta. The Bloch Hamiltonian will be constructed in terms of  $h_0(\bt{k})$ which is a single layer component with no twist local (in $k$-space) to the Dirac node centered at $K_N$ expressed as
\begin{equation}
    h_0(\bt{k}) = \xi \tau_3s_0 + \delta \hat{\Delta}(\bt{k}),
    \label{eq:node_h0}
\end{equation} 
where $\xi = v_F k_\parallel$, $\delta = v_\Delta k_\perp$, $\tau_i$ is the Nambu basis, and $s_i$ is the spin basis. The parameters $v_F$ is the velocity of the normal dispersion and $v_\Delta$ is the ``velocity'' of the order parameter, and $k_\parallel$  is the momenta parallel to $K_N$ and  $k_\perp$ is the momenta perpendicular to $K_N$. Local to the node at $\bt{K}_N$, $k_\parallel$ and $k_\perp$ forms an orthogonal coordinate system (see Fig.~\ref{fig:coords}). For our purposes, we consider the order parameter $\hat{\Delta}(\bt{k})$ to be expressed as  
\begin{equation}
    \hat{\Delta}_s = s_2 \tau_2,
    \label{eq:OP_singlet}
\end{equation} 
or 
\begin{equation}
    \hat{\Delta}_t(\bt{k}) = ({\bt{d}(\bt{K}_N)}\cdot \bt{s}) (i s_2)\tau_1,
    \label{eq:OP_triplet}
\end{equation} 
which are singlet and  triplet order parameters, respectively. 
Without loss of generality, we'll chose the $d$-vector of the form $\bt{d}(\bt{K}_N) = ( d_1(\bt{K}_N),0,0)$ where $ d_1(\bt{K}_N) = -d_1(-\bt{K}_N)$ and $|d_1(\bt{K}_N)| = 1 $.

Treating the tunneling only between nearest neighboring layers and identical between each layer, a general untwisted $N$ layer Hamiltonian can be written as
\begin{equation}
    H_N(\bt{k}) = 
    \begin{pmatrix}
    h_0(\bt{k})   & T(\bt{k})                       & 0 &  \\
    T^*(\bt{k})                        & h_0(\bt{k}) & T(\bt{k}) &  \dots \\
    0                                    & T^*(\bt{k})                           & h_0(\bt{k}) &  \\
                                        & \vdots                             &  & \ddots 
\end{pmatrix}   
\label{eq:Ham_Nlayers}
\end{equation}

%To examine the topological nature of twisted superconductors,
\noindent where the off-diagonal interlayer tunneling matrix $T(\bt{k})$ is represented in terms of bases $\tau_i$ and $s_j$.  We'll now address the affect of introducing a small twist of angle $\pm\theta$ to an arbitrary layer $n$. The twist translates this layer's node by introducing a displacement $\bt{Q}_N$ along $k_\perp$ as shown in Fig.~\ref{fig:coords} (for more details see Ref.~\cite{Volkov2022_2}). This affects the tunneling matrix by involving both the untwisted momentum components $\bt{k}$ and twisted momentum ${\bf k}'={\bt{k}}^\theta$ such that $T({\bf k})\rightarrow T({\bf k},{\bf k}')\approx T({\bf k}-{\bf k}')$ and
\begin{equation}
    T(\bt{k}-{\bt{k}}') = t_{\bt{k},{\bt{k}}'} \tau_i s_j
\end{equation}
where $t_{\bt{k},{\bt{k}}'}$ is the Fourier transformed tunneling strength. Treating the tunneling strength as local in $k$-space to the node $\bt{K}_N$, and neglecting momenta beyond the initial Brillouin zone as shown in Fig.~\ref{fig:coords} as $T$ falls off exponentially in this regime~\cite{Volkov2022_2}, the tunneling strength can therefore be simplified to 
\begin{equation}
    t_{\bt{k},{\bt{k}}'} \approx \frac{t_{\bt{K}_N}}{\Omega} \equiv t
\end{equation}
where $t_{\bt{K}_N}$ is the tunneling strength at the node $\bt{K}_N$ and $\Omega$ is the unit cell area of a layer (for technical details of this approximation, see Ref.~\cite{Volkov2022_2}).
Furthermore, the displacement also contributes a term $H_{tw,n}(\bt{k})$ which is added to the twisted layer's Hamiltonian of Eq.~\eqref{eq:node_h0}
\begin{equation}
    H_{tw,n}(\bt{k}) = \pm \alpha t \hat{\Delta}(\bt{k}),
    \label{eq:twist_H}
\end{equation}
where $ \alpha = \frac{v_\Delta Q_N}{t}$ and the subscript $n$ is the layer index where the twist is applied.

We now aim to classify the Hamiltonian under the discrete symmetry operators for time-reversal $\mathcal{T} = U_T K$ and charge-conjugation $\mathcal{C} = U_C K$ where $U_{T,C}$ are unitary operators. As the contributions of twist do not affect the classification of the Hamiltonian in Eq.~\eqref{eq:Ham_Nlayers}, they are neglected. These operators realize the $\mathcal{T}$ and $\mathcal{C}$ symmetries of the Hamiltonian as: 
%act on the Bloch Hamiltonian as \pv{[PV: you haven't specified whether there is spin or not; with spin I don't think it's correct, i.e. $s_z\to=-s_z$ under TRS]} \kl{KL: This is true regardless of spin, spin changes the form of the unitary operator to account for this sign change. i.e. for TRS, no spin means $U_T = \mathbb I$ but with spin $U_T = i s_2 $ }
%
\begin{eqnarray}
    U_T^\dagger H_N^*(\bt{k}) U_T = H_N(-\bt{k}), \,\, U_C^\dagger H_N^*(\bt{k}) U_C = -H_N(-\bt{k}).
    \label{eqn:symmetries}
\end{eqnarray}
\noindent It's important to note that $H_N(\bt{k})$ represents a single valley centered at $ \bt{K}_N$ and these discrete operations relate this valley to another located at $-\bt{K}_N$.  Since $H_N(\bt{k})$ can be decomposed into  single layer components, we can likewise decompose the unitary operators $U_T$ and $U_C$ in Eq.~\eqref{eqn:symmetries} as a direct sum

\begin{equation}
    U_{T/C} = U^{(1)}_{1,T/C} \oplus U^{(2)}_{1,T/C} \oplus \dots \oplus U^{(N)}_{1,T/C},
\end{equation}
where the superscript corresponds to the layer index and $U_{1,T/C}$ is the unitary operator for time-reversal/charge-conjugation corresponding to the single layer Hamiltonian such that
\begin{eqnarray*}
    U_{1,T}^\dagger h_0^*(\bt{k}) U_{1,T} = h_0(-\bt{k}), \\ 
    U_{1,C}^\dagger h_0^*(\bt{k}) U_{1,C} = -h_0(-\bt{k}).
\end{eqnarray*}
These operators will also act on the tunneling matrix which must obey the same relation above. We now focus on the spin degrees of freedom as they play a crucial role in the form of the unitary operators and ultimate topological classification of the superconductor. We start by considering the unitary operators action on $h_0(\bt{k})$. With only a singlet component ($\hat{\Delta} = \hat{ \Delta }_s$), the Hamiltonian  will have full spin $SU(2)$ symmetry, corresponding to a CI class with the time-reversal symmetry operator $\mathcal{T} = K$ and charge-conjugation given by $\mathcal{C} = \tau_2 K $. For a triplet superconductor ($\hat{\Delta} = \hat{ \Delta }_t$), the $d$-vector defining the triplet order parameter characterizes the spin degrees of freedom. For our purposes, we'll assume we  the $d$-vector has a single component that breaks the $SU(2)$ symmetry down into $U(1)$ and leaves us in class AIII with time-reversal $\mathcal{T} = is_2 K$ and charge-conjugation $\mathcal{C}= \tau_1 K$. 

We'll now include perturbations to this Hamiltonian, some of which can be generated by an 
external field or an applied current (see below) that can be added to either $h_0(\bt{k})$ or $T(\bt{k})$  
of the form
\begin{comment}
\begin{equation}
    H_{\mathrm{pert}}^{\alpha\beta}(\bt{k}) = \sigma_\alpha\tau_\beta(\bt{h}(\bt{k})\cdot\bt{s} ),
\end{equation} 
where $\alpha,\beta=0,1,2,3$
\end{comment}
\begin{equation}
    H_{\mathrm{pert}}^{\alpha\beta}(\bt{k}) = \tau_\alpha({h}_\beta(\bt{k})s_\beta ),
\end{equation} 
where $\alpha,\beta=0,1,2,3$ and $\bt{h}(\bt{k})$ is a spin and momentum dependent parameter corresponding to an external field. For our purposes we consider each term independently, and we aim to find perturbations which 
\begin{enumerate}[(i)]
\item preserve $\mathcal{T}$ and $\mathcal{C}$ in Eq.~\eqref{eqn:symmetries},
\item open a gap in the spectrum,
\item have a nontrivial topological index.
\end{enumerate}
\noindent To satisfy condition $(i)$ for a superconductor, we are automatically restricted to class DIII to form a $\mathbb{Z}_2$ invariant. Analyzing the possible perturbations under the discrete symmetries (see Appendix \ref{appendix:A}) leads to the term 
%$
\begin{equation}
H_{JSC}(\bt{k}) =  h(\bt{k})\sigma_0\tau_2s_0
\label{eq:HJSC}
\end{equation}
%$ 
that satisfies conditions $(i)$ and $(ii)$ so long as $h(\bt{k}) = - h(-\bt{k})$.   Such a term can be considered as adding an $i p$ component to the order parameter with $d$-vector component along $s_2$.

%A key figure of this term is that it depends on a triplet order parameter in order to satisfy Fermi-Dirac statistics by being spatially odd but spin even. 
Importantly, not all the momentum-odd character of Eq. \eqref{eq:HJSC}
is consistent with Fermi-Dirac statistics. The effect of which results in differences for singlet and triplet superconductors. For a triplet superconductor, this term produces a helical order parameter while for singlet superconductors it violates Fermi-Dirac statistics. 
%Then 
To amend this for singlet superconductors, such a term would require introducing a triplet order parameter which would produce a class AIII superconductor with a trivial topological classification in 2D~\cite{Schnyder2008}. The affect on a triplet order parameter, however, is allowed and sufficient to produce a class DIII superconductor. Starting with the two layer case, we show how such a component is generated by a Josephson spin current in triplet superconductors. To then satisfy condition $(iii)$, we will compute a $\mathbb Z_2$ index which will be shown to depend on $N$ and twist arrangement.

%

%%%%%%%%%%%%
%%%%%%%%%%%%
\section{Topological Gap Opening With Josephson Spin Current}
%%%%%%%%%%%%
%%%%%%%%%%%%
\label{sec:JSC}

Before considering the stack of $N$ layers, we first consider the effect of $H_{JSC}(\bt{k})$ defined in Eq.~\eqref{eq:HJSC} on the low-energy spectrum of a twisted bilayer:
\begin{equation}
    H_2(\bt{k}) = \xi \tau_3 + \delta d_1(\bt{K}_N) \tau_1 s_3  -\alpha t d_1(\bt{K}_N) \tau_1 \sigma_3 s_3 + t \tau_3 \sigma_1 , 
    \label{eq:bi_Ham0}
\end{equation}
where we take a triplet order parameter, and $\sigma_i$ corresponds to the layer basis. Since all three rotation arrangements are equivalent for two layers, we take the layer to be rotated in opposing directions (incorporated in the $\sigma_3$ term) by a small angle $\pm\frac{\theta}{2}$. Compared to Eq.~\eqref{eq:twist_H}, the parameter $\alpha$ is halved due to half the rotation angle applied. The off-diagonal tunneling matrix is also captured by $\sigma_1$ and is given an explicit form of $T(\bt{k}) = t \tau_3$ which addresses the essential physical affects of twist while simplifying it's form~\cite{Volkov2022_2}.

\begin{figure}[h]
\centering
    \includegraphics[width=\columnwidth]{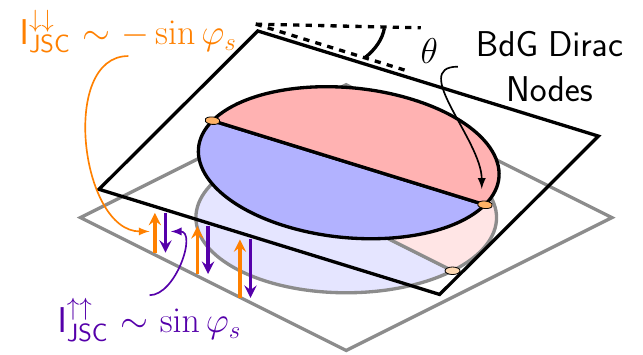}
    \caption{Cartoon depiction a two nodal $p$ wave superconductors rotated by an angle $\theta$. The location of the Dirac nodes are denoted by orange dots. Interlayer spin current ($I^{\sigma\sigma}_{JSC}$ where $\sigma$ refers to the spin channel) results in a spin Josephson effect. Consequently, a rotation in spin space of the d-vector and Nambu space of the order parameter occurs which forms a $p_x+i p_y/ p_x - i p_y$ order parameter.
    }
        
    \label{fig:2layer_JSC}
\end{figure}

%In this approach, an interlayer current is applied which induces a phase difference between the layers. 
The application of the conventional charge (Josephson) supercurrent between layers induces a phase difference $\varphi$ between the layers. This generates a term $\propto\tau_2$ in the low-energy Hamiltonian that opens a gap at the Dirac nodes and breaks time reversal symmetry~\cite{Volkov2022,Volkov2022_2}.
Using the concept of the interlayer current, we can relate $H_{JSC}(\bt{k})$ to an interlayer spin current illustrated in Fig.~\ref{fig:2layer_JSC}. For triplet superconductors, spin currents have been studied in junction devices with two triplet superconducting layers with a material separation. In these arrangements, spin currents are induced by misalignment of the $d$-vector of the order parameter~\cite{Asano2006,Rashedi2007} or with the magnetic moments of a magnetic separation layer~\cite{Brydon2009}. Here, we describe the spin current analogously to an interlayer current where the order parameter will acquire a spin-phase, %except 
i.e. with opposing signs for each spin channel such that $ \Delta_1 \rightarrow \Delta_1 \big( e^{i\varphi_s/2}s^{\uparrow \uparrow} + e^{-i\varphi_s/2}s^{\downarrow\downarrow} \big)$ and $ \Delta_2 \rightarrow \Delta_2 \big( e^{-i\varphi_s/2}s^{\uparrow \uparrow} - e^{i\varphi_s/2}s^{\downarrow\downarrow} \big)$ where $s^{\uparrow \uparrow} = \frac{s_0 + s_3}{2}$ and  $s^{\downarrow \downarrow} = \frac{s_0 - s_3}{2}$. By using the current-phase relationship, we can represent the current of each spin channel as $I^{\sigma\sigma}(\Delta \varphi_s)$ where $\Delta\varphi_s$ is the spin-phase difference between the layers. 
The Josephson (spin) current $I_{JC}$ ($I_{JSC}$) is then the sum (difference) of the spin channels~\cite{Mao2022}. In the context of a Josephson junction, an applied spin current will induce a spin-dependent phase difference ($\varphi_s$) that generates a Josephson spin current. This realizes the spin Josephson effect, which is given by (see Appendix \ref{appendix:josephson})
\begin{equation}
    I_{JSC}=\frac{2t^2 \nu_0 }{\hbar}\sin(\varphi_s),
\end{equation}
where $\nu_0$ is the density of states at the Fermi energy. Since the phase difference for each spin channel is equal and opposite, the current (spin current) will be zero (non-zero). For a triplet superconductor, this acquired phase results in the transformation of the $d$-vector to $\bt{d}(\bt{K}_N) = d_1(\bt{K}_N)(\cos \frac{\varphi_s}{2},\sin \frac{\varphi_s}{2},0)$, resulting in the effective low-energy Hamiltonian 
$H_2(\bt{k}) \rightarrow \tilde{H}_{2}(\bt{k}) + H_{2,SJC}(\bt{k})$  where
\begin{eqnarray}
     \tilde{H}_{2}(\bt{k}) & = & \xi \tau_3 + \delta \cos\frac{\varphi_s}{2}d_1(\bt{K}_N)\tau_1s_3  \nonumber \\
    & - & \alpha t \cos\frac{\varphi_s}{2}d_1(\bt{K}_N)\sigma_3\tau_1s_3 + t \sigma_1 \tau_3 
    \label{eq:trip_Ham1}
\end{eqnarray}
and
\begin{equation}
     H_{2,SJC}(\bt{k}) =  -\delta \sin\frac{\varphi_s}{2} d_1(\bt{K}_N) \sigma_3\tau_2  + \alpha t \sin\frac{\varphi_s}{2}d_1(\bt{K}_N)\tau_2 
    \label{eq:trip_HamSJC}
\end{equation}
\noindent Note the latter term in $H_{2,SJC}(\bt{k}) $ corresponds to the gap-opening perturbation identified in Sec.~\ref{sec:sym}. Compared to the interlayer current, this spin current converts the Hamiltonian into a class DIII topological superconductor with helical edge modes of the form $p_x + i p_y/p_x - i p_y$. For singlet superconductors, the proposed mechanism of generating a TRI gap would not work, as the singlet order is even under ${\bf k}\to-{\bf k}$. However, the fermionic commutation relations imply that $H_{JSC}({\bf k})$ has to be odd in ${\bf k}$, or, in other words, will vanish identically if it's assumed to be even in ${\bf k}$. Moreover, applying a spin phase difference to an even order parameter can only generate even terms, and thus the term of interest will not be generated. Therefore, opening a TRI gap in singlet SC would require the formation of a secondary spin triplet order parameter. %\kl{KL:apparently mixed order parameters are permitted when a system lacks inversion. I need to review citations again.} 
Furthermore, if such a term is induced, it is insufficient  to produce a stable $\mathbb{Z}_2$ invariant which is explored further in Appendix~\ref{appendix:singlet} and noted in Sec.~\ref{sec:sym}. Returning to the triplet scenario, if we take the spin current as small such that $\varphi_s \ll 1$, we can treat $H_{2,SJC}(\bt{k})$ as a perturbation. The former term vanishes to first order and produces a negligible term of $\mathcal{O}(\varphi_s^2)$. Projected Eq.~\eqref{eq:bi_Ham0} and the latter term of Eq.~\eqref{eq:trip_HamSJC}  into the zero energy basis of Eq.~\eqref{eq:bi_Ham0}, we find the effective Hamiltonian
%
\begin{comment}
\begin{equation}
\mathcal{H}_{low}(\bt{k}) = \begin{pmatrix}
    -\tilde{\xi} & -\tilde{\delta} + i \frac{t\alpha}{2} \varphi_s  & 0 & 0 \\
    -\tilde{\delta} - i \frac{t\alpha}{2} \varphi_s  & \tilde{\xi} & 0 & 0 \\
    0 & 0  & \tilde{\xi} & \tilde{\delta} - i \frac{t\alpha}{2} \varphi_s  \\
    0 & 0  & \tilde{\delta} + i \frac{t\alpha}{2} \varphi_s  & -\tilde{\xi}
\end{pmatrix}
\end{equation}
\end{comment}
\begin{equation}
    \mathcal{H}_{\mathrm{low}}(\bt{k}) = -\tilde{\xi} \zeta_3\eta_3 - \tilde{\delta} \zeta_1\eta_3 - \tilde{m} \zeta_2\eta_0
\end{equation}
where $\zeta_i$ and $\eta_j$ form our low energy basis, $\tilde{\xi} = \sqrt{1-\alpha^2}\xi$, $\tilde{\delta} = \sqrt{1-\alpha^2} \delta$, and $\tilde{m} = \frac{t\alpha}{2} \varphi_s$. Each $2$x$2$ block acts as  a $p_x \pm i p_y$ superconductor which provide the helical counter-propagating edge modes.

We now aim to quantify the topological invariants of the Hamiltonian. A general approach is taken where we distinguish the eigenstates by diagonalizing them in the degenerate subspace with the $\hat{S}_z$ operator. For our two occupied states $|\nu_\alpha \rangle$ where $\alpha = \pm$, this amounts to diagonalizing the matrix
\begin{equation}
    S_{\alpha,\beta} = | \nu_\alpha(k) \rangle\langle \nu_\alpha(k) | \sigma_z | \nu_\beta(k) \rangle \langle \nu_\beta(k) |.
\end{equation}
The Chern number for the occupied eigenstates of the matrix can be computed as
\begin{equation}
C_{\pm} = \frac{1}{2\pi}\int d^2 k  Q_{K\pm}(k)
\end{equation}
\noindent where~\cite{Yang2011} 
\begin{equation}
    Q_{K\pm}(k) = i \Big( \langle \partial_{k_\parallel} \nu_\pm | \partial_{k_\perp} \nu_\pm \rangle - \langle \partial_{k_\perp} \nu_\pm | \partial_{k_\parallel} \nu_\pm \rangle \Big).
\end{equation} 
Performing this calculation, we find $C_\pm = \pm \frac{\textrm{sign}( t\alpha\varphi_s) }{2}$. With these terms, we can recognize the spin Chern number $\mathcal{C}_s$
\begin{equation}
    \mathcal{C}_s = C_+ - C_- = \textrm{sign}( t\alpha\varphi_s ),
    \label{eq:spin_chern_form}
\end{equation}
while the Chern number $\mathcal{C}$ is
\begin{equation}
    \mathcal{C} = C_+ + C_- = 0. 
\end{equation}
The $\mathbb{Z}_2$ invariant can then be computed as the total spin Chern number by summing over all layers as each one contributes an additional node

\begin{eqnarray}
 \nu  &=& \sum^N_i \mathcal{C}_s  \mod 2 \equiv \mathcal{C}_{s,tot}  \mod 2.
% \\ 
%&=& N \,\mod 2.
\label{eq:tot_chern}
\end{eqnarray}
For the case of the two layer Hamiltonian, this leads to a trivial $\mathbb{Z}_2$ index always, as the numbers of valleys, each containing two Dirac points, is even. Assuming each layer contributes a non-zero $\mathcal{C}_s$, we can naively expand this result to any layer number $N$ where any even number of layers is topologically trivial, but opens the possibility of a non-trivial $\mathbb{Z}_2$ index for an odd number of layers. The complication with this general argument is that the twisting arrangements for more than two layers is not unique (see Fig.~\ref{fig:twisting_arrangements}) and may affect the topology of the system. Using three layers as an example, we will consider several simple twisting arrangements to show when Eq.~\eqref{eq:tot_chern} can be applied to any $N$ layered system. In particular, we arrive at a non-trivial $\mathbb Z_2$ invariant for odd layered systems for CT and SLT set ups. We are then able to extend this result to $N$-layers for the SLT cases.

%%%%%%%%%%%%%%%
%%%%%%%%%%%%%%%
\section{Trilayer Twists}
%%%%%%%%%%%%%%%
%%%%%%%%%%%%%%%
\label{sec:3layer_twists}

Given a spin current, the gap that is introduced provides a well-defined spin Chern number for a node. Although $\nu = 0$ for bilayers, an additional node in a trilayer system can provide $\nu =1$ so long as the additional node becomes gapped. To express the Hamiltonian, we'll start with Eq.~\eqref{eq:Ham_Nlayers} for $N=3$ and assume the tunneling matrix $T = t \tau_3$ as in Sec.~\ref{sec:JSC} resulting in
\begin{equation}
    H_{3}(\bt{k}) = \begin{pmatrix}
    h_0(\bt{k})   & t \tau_3 s_0                        & 0 \\
    t \tau_3s_0                            & h_0(\bt{k}) & t\tau_3s_0 \\
    0                                    & t \tau_3s_0                             & h_0(\bt{k})
\end{pmatrix} .
\end{equation}

\begin{figure}
\centering
         \includegraphics[width=.4\textwidth]{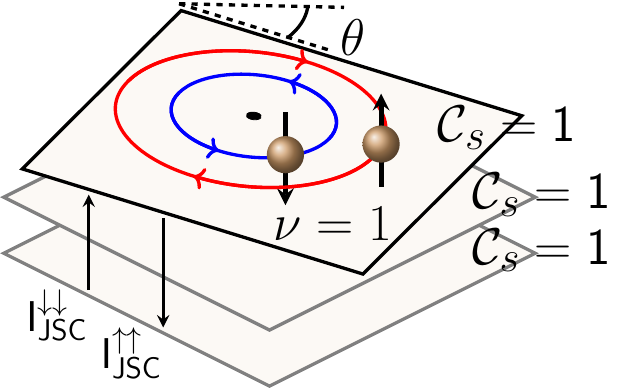}
%    \caption{$\mathbb{Z}_2$ index for trilayer system. We apply a twist to the top layer of angle $\theta$ and an equal spin interlayer current. Here $I^{\sigma,\sigma}$ represents the current spin channel of the order parameter. When the spin currents are polarized in opposing directions, a Dirac mass is introduced in the model which forms a spin Chern number for each layer. The BdG quasiparticles in orange can be given a helicity represented by the brown arrows. These quasiparticles will form helical edge modes around impurities in the lattice. In the case of three layers, a stable $\mathbb{Z}_2$ index of $\nu = 1$ can be formed by summing the spin Chern number of each layer given by $C_s$.}
    \caption{Edge modes due to a finite $\mathbb{Z}_2$ index for the SLT trilayer system. For a SLT trilayer system with a interlayer spin current, a Dirac mass is introduced which forms a spin Chern number for each layer. Here $I^{\sigma,\sigma}$ represents the current spin channel of the order parameter. The BdG quasiparticles in brown can be given a helicity represented by the black arrows. These quasiparticles will form helical edge modes around impurities in the lattice. The corresponding $\mathbb{Z}_2$ formed by  summing the spin Chern number provides a nontrivial index $\nu = 1$.
    }
    \label{fig:ILSSC_cartoon}
\end{figure}
\noindent To introduce twists, we need to add the angle dependent terms of Eq.~\eqref{eq:twist_H} corresponding to each twisting arrangement. Following the twisting orientations of Fig.~\ref{fig:twisting_arrangements}, alternating twist (AT) will have twists applied in opposing orientations which leads to a twisting Hamiltonian 
\begin{equation}
    H_{tw,AT}(\bt{k}) = \sum_{n=1}^N  (-1)^n \alpha  d_1(\bt{K}_N) \tau_1 s_1  \delta_{n,n},
\end{equation}
where $\delta_{n,n}$ is the Kronecker delta acting in layer space.
Chiral twisting (CT) instead have the twist angle in the same orientation leading to a contribution
\begin{equation}
    H_{tw,CT}(\bt{k}) = \sum_{n=1}^N  n \alpha  d_1(\bt{K}_N) \tau_1 s_1  \delta_{n,n}.
\end{equation}
For single layer twisting (SLT), only a boundary layer is rotated while the other layers remain fixed which introduces a twisting Hamiltonian 
\begin{equation}
    H_{tw,SLT}(\bt{k}) = \alpha  d_1(\bt{K}_N) \tau_1 s_1  \delta_{n,1}
\end{equation}
where we chose the first layer as the rotated layer.

\begin{figure}
    \centering
         \includegraphics[width=.45\textwidth]{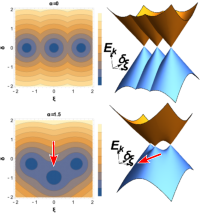}
    \caption{Dispersion of the SLT Hamiltonian. For  no twist ($\alpha = 0$) , three nodes lie along $\xi$. As twist starts to increase, the nodes move together briefly. Afterwards, as shown when $\alpha =1.5$, the central node moves away from the other two nodes along $\delta$ while the other nodes remain fixed. Red arrow depicts the direction the central node follows as $\alpha$ increases.
    }
    \label{fig:SLT_dispersion}
\end{figure}

Each case behaves differently as $\alpha$ increases. For the case of ATs, the system can be treated as a bilayer and decoupled monolayer due to their different behavior under a mirror plane parallel to the layers~\cite{Khalaf2019,classen2022interaction,Tummuru2022}. As a result, an interlayer supercurrent only induces a gap for the coupled bilayer while the monolayer's nodes remain fixed. For CTs and SLTs, all layers are coupled where an interlayer supercurrent can gap all nodes in the layers producing a topological phase. One difference between the SLTs and the other two is that only ATs and CTs can achieve a magic angle at $\alpha_{MA} =\sqrt{2}$ marked by a quadratic and cubic band touching, respectively~\cite{Tummuru2022}. However, as Figure \ref{fig:SLT_dispersion} displays, SLTs form no band touching. Instead, at $\alpha = 0$, the nodes lie along $\delta = 0$ and as $\alpha$ starts to increase briefly causes the nodes along $k_\parallel$ to converge towards the center node. The node at the center then travels along $k_\perp$ away from the two nodes which remain fixed along $k_\parallel$. Further increasing $\alpha$ causes the single node to continue to travel further away along $k_\perp$.

\begin{figure}
\centering
         \includegraphics[width=.45\textwidth,center]{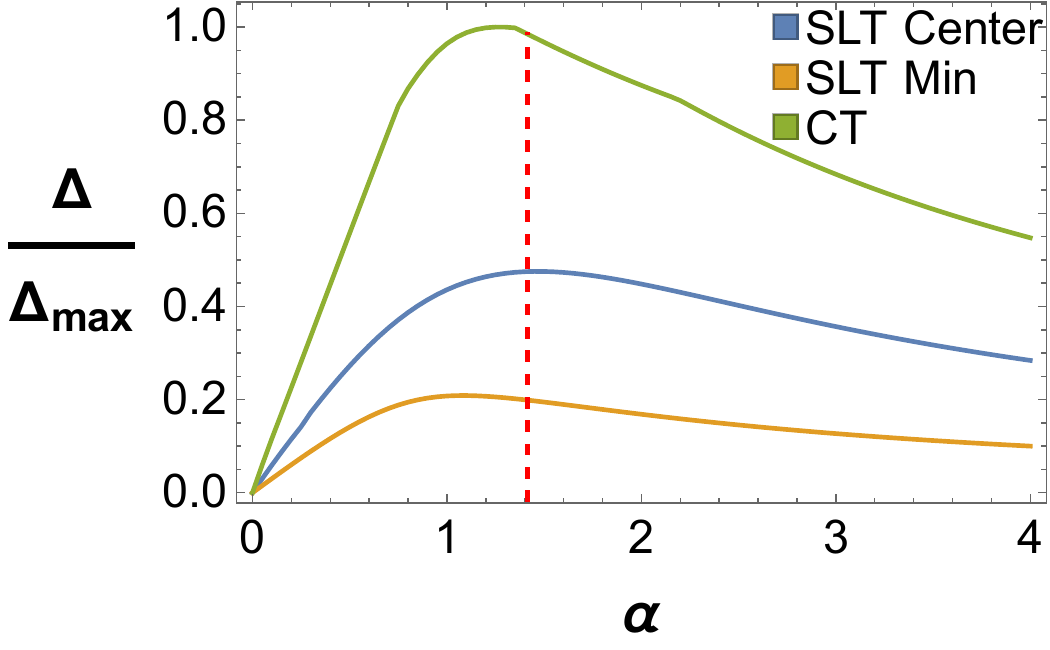}
    \caption{Normalized gap energy versus $\alpha$. For CTs, all nodes have the same gap which plateaus at $\alpha_{MA} = \sqrt{2}$ highlighted by the red line. For SLTs, all nodes are gapped but with different magnitude. This difference is shown by the plots of the gap from the central node in blue and the minimum gap contributed by the remaining nodes in orange. Here, the gap energy is normalized by the maximum gap energy $\Delta_{max}$, which is taken from the CTs' nodes.
    }
    %With the application of a spin current, a gap opens in the dispersion for all $\alpha >1$. Plotted are the gap for the CTs and SLTs trilayers as $\alpha$ increases. For the CTs, all nodes have the same energy gap. The gap energy increases until  $\alpha=2$ after which decreases for increasing $\alpha$. For SLTs, the single node which separates has a different gap to the pair of nodes which remain along $\delta =0$. 
    \label{fig:twist_gap}
\end{figure}

The results for both CTs and SLTs have a convoluted analytical form, but we can obtain their response to an interlayer spin current numerically. Once a spin polarized spin difference is applied, the dispersion of both twist arrangements becomes gapped for $\alpha \neq 0$. Figure \ref{fig:twist_gap} shows how the gap energy evolves under increasing twist angle via $\alpha$. For CTs, all nodes have the same gap energy which increases until $\alpha_{MA} =\sqrt{2}$ where it plateaus and then decreases as $\alpha > \sqrt{2} $. For SLTs, the gap energy is different between the center node and the pair of nodes along $\xi$. 
As $\alpha$ increases, the central node's gap energy exceeds the gap of the other two nodes. Regardless, the gaps follow the trajectory of the CTs' nodes albeit with smaller magnitude for all $\alpha > 0$.

To determine the topological invariant formed from the interlayer spin current, we focus on the SLT case where we can treat the twist angle as small such that $\alpha \ll 1$. In this approximation, we start with an untwisted stack where the nodes lie along $K_N$. In doing so, the contribution of the order parameter  and the twist along $k_\perp$ are perturbations about the zero energy eigenstates. The untwisted Hamiltonian can be represented by $\mathcal{H}_{SLT}(\bt{k})$
\begin{equation}
    \mathcal{H}_{SLT}(\bt{k})   = \sum_{l=1}^{N=3}  \Psi^\dagger_{\bt{k},l} \xi \tau_3 \Psi_{\bt{k},l} + \left(\Psi^\dagger_{\bt{k},l-1} t \tau_3 \Psi_{\bt{k},l} + \mathrm{h.c.}\right)
    %\nonumber \\
    %&+ \Psi^\dagger_{\bt{k},l+1} t \tau_3 \Psi_{\bt{k},l},
\end{equation} 
where $\Psi^\dagger_{\bt{k},l}$ are the same as discussed in Eq.~\eqref{eq:H_general}. The zero energy eigenstates for this Hamiltonian can be written as a superposition of Bloch states using open boundary conditions for the layer hopping,
\begin{equation}
      | \Psi^\dagger_{k_i^z} \rangle  = \frac{1}{2}\sum_{l=1}^{N=3} \sin (k_i^z l) \Phi^\dagger_{\bt{k},l} | 0 \rangle,
\end{equation}
where $\Phi^\dagger_{\bt{k},l}$ is a B-W spinor composed of $ \Phi^\dagger_{\bt{k},l} = \Psi^\dagger_{\bt{k},l} - \Psi^\dagger_{\bt{k},-l}$. The momenta $k_i^z = \frac{i\pi}{4},\; i = 1,2,3$ is oriented along the $z$-direction which points perpendicular to the $(k_\parallel, k_\perp)$ plane along the stack of layers. The discrete index $i$ comes from Fourier transforming the layers and therefore corresponds to the number of layers in the system.

We'll now introduce a dispersion along $k_\perp$ which introduces a contribution from the order parameter as well as the interlayer spin current,

\begin{equation}
    \mathcal{H}_\delta(\bt{k}) = \delta \sum_l \Psi^\dagger_{\bt{k},l} \left( \cos(\varphi_s l ) \tau_1s_3 -  \sin(\varphi_s l ) \tau_2s_0 \right) \Psi_{\bt{k},l},   
\end{equation}
such that each layer has a phase $e^{i\varphi_s l}s^{\uparrow\uparrow} + e^{-i\varphi_s l}s^{\downarrow\downarrow} $ which ensures a phase difference between layers and opposing current in the spin channels. Applying the twist to the first layer  adds a perturbation along the perpendicular direction,
\begin{equation}
    \mathcal{H}_{tw}(\bt{k},\varphi_s) = \alpha t \Psi^\dagger_{\bt{k},1} \left( \cos\varphi_s \tau_1s_3 + \sin\varphi_s \tau_2s_0 \right) \Psi_{\bt{k},1},
\end{equation}
where the phase is incorporated from the spin current. Projecting these perturbations onto our low energy basis, we find %\pv{[PV: Why does the effect vanish at $\varphi_s = \pi/2$?]} \kl{[KL: this is when new nodes appear, I added a paragraph to address and direct this affect to the other article.] }
\begin{equation}
    \mathcal{H}_\delta(k_i^z) = \sum\limits_{k_i^z}  z_\delta(k_i^z) \Psi^\dagger_{k_i^z}\Big( \cos(2 \varphi_s) \tau_1s_3 - \sin(2 \varphi_s) \tau_2s_0 \Big)  \Psi_{k_i^z}
\end{equation}
where 
%
%\begin{equation}
% z_\delta(k_i^z)  = 2\delta\frac{\cos^2 \frac{\varphi_s}{2}\cos \varphi_s}{\cos \varphi_s - \cos(2k_i^z)}\sin^2 k_i^z 
%\end{equation}
%
\begin{equation}
 z_\delta(k_i^z)  = \delta\frac{\cos^2\frac{\varphi_s}{2}\cos\varphi_s}{2\sin(k_z^i-\frac{\varphi_s}{2})\sin(k_z^i+\frac{\varphi_s}{2}) }\sin^2 k_i^z,
\end{equation}
and
\begin{equation}
    \mathcal{H}_{tw}(k_i^z) = \sum_{k_i^z}  z_{tw}(k_i^z) \Psi^\dagger_{k_i^z}\Big( \cos\varphi_s \tau_1s_3 - \sin\varphi_s \tau_2s_0 \Big)  \Psi_{k_i^z},
\end{equation}
where 
\begin{equation}
 z_{tw} (k_i^z) = \frac{\alpha t}{2} \sin^2 k_i^z. 
\end{equation}
The low energy effective Hamiltonian about each node then becomes
\begin{equation}
    \mathcal{H}_{\mathrm{eff}}(k_i^z) = \sum\limits_{k_i^z} \Psi^\dagger_{k_i^z}\begin{pmatrix}
        \tilde{\xi} s_0  & \tilde{\delta} s_3 - i\tilde{m}s_0 \\
        \tilde{\delta}s_3 + i\tilde{m}s_0  & -\tilde{\xi} s_0 
    \end{pmatrix}\Psi_{k_i^z},
\end{equation}
where $\tilde{\xi} =\xi- \xi_{k_i^z}$, $\tilde{\delta} = z_\delta(k_i^z)\cos(2\varphi_s) +  z_{tw}(k_i^z) \cos \varphi_s$ and $\tilde{m} = z_\delta(k_i^z)\sin(2\varphi_s) + z_{tw}(k_i^z) \sin \varphi_s$. 

Since the contribution of both twist and current to the Dirac mass is always $|\tilde{m}| >0$, this leads to a well-defined Chern number. Furthermore, since $\tilde{\delta} >0$ for all $k_i^z$, all gapped Dirac nodes will produce a Chern number with the same sign. Since the corresponding occupied eigenstates are orthogonal eigenstates of $\hat{S}_z$, we can evaluate the spin Chern number following Eq.~\eqref{eq:spin_chern_form} such that each node contributes 
\begin{equation}
    \mathcal{C}_s = \textrm{sign}( \alpha t \sin \varphi_s ).
\end{equation} 
For this three layer case, we find
$\mathcal{C}^{(3)}_{s,tot}\; \textrm{mod}\;2 = 1$, leading to a stable $\mathbb{Z}_2$ index. 

For SLTs we can now generalize this approach to $N$ layers, allowing us to write the $\mathbb Z_2$ index as
\begin{eqnarray}
\nu = \mathcal{C}^{(N)}_{s,tot}\; \textrm{mod}\;2 = \sum_{l=1}^N \mathcal{C}_s  \; \textrm{mod}\;2 \nonumber \\
= \begin{cases}
    0, \quad N \in 2\mathbb{N} \\
    1, \quad N \in 2\mathbb{N} + 1
\end{cases}.
\label{eq:z2_inv}
\end{eqnarray}
Thus, for the SLT geometry we are able to complete the computation of the $\mathbb{Z}_2$ index  for $N$-layers spelled out in Eq.~\eqref{eq:tot_chern}.
On closer inspection of $z_\delta(k_i^z)$, when specifically choosing $\varphi_s = \frac{\pi}{2}$, this quantity vanishes. This result consequently marks a transition to new Dirac nodes appearing along $k_\perp$. Examined numerically, a pair of nodes appear at $k_2^z = \frac{\pi}{2}$ along $k_\perp$ leading to a total of three nodes at $k_2^z$. A more detailed investigation of current-driven generation of new Dirac nodes is provided in Ref.~\cite{lucht2023}. One consideration for these newly generated nodes is if they affect the $\mathbb{Z}_2$ invariant once formed. Because these new nodes are gapped and always introduced as pairs, their contribution will cancel and will have no affect on the $\mathbb{Z}_2$ invariant. Therefore, Eq.~\eqref{eq:z2_inv} can be applied to any finite odd layered system and provide a nontrivial $\mathbb{Z}_2$ index.

%\kl{ The possible influence of these nodes on the Chern number, and their appearance for varying layer number and $\varphi_s$ is explored in detail in the accompanying article.}
%
%

%%%%%%%%%%%%%%%
%%%%%%%%%%%%%%%
\section{Discussion and Conclusion}
%%%%%%%%%%%%%%%
%%%%%%%%%%%%%%%
\label{sec:fin}

For $N$ layered twisted nodal superconductors, we demonstrate how twist and spin current can produce a TRI topological phase. Starting with the AZ classifications on perturbations on a $N$-layer system, we identify one which is related to a Josephson spin current. With the Josephson spin current, we show that triplet twisted nodal superconductors transform its AZ classification to a DIII superconductor with a well-defined spin Chern number. This spin Chern number corresponds to a $\mathbb Z_2$ topological invariant which can be non-trivial for odd layered systems depending on the twisting arrangement, which is not unique. 

To determine the influence of the twisting arrangement, we study a three layer system under common twisting configurations to show how chiral and single layer twisting arrangements form a topological gap with an interlayer spin current. For the single layer twists, we show with an effective low energy model that this topological gap is accompanied by a non-trivial $\mathbb Z_2$ invariant. Extending into higher layer numbers where 2.5-D affects modify the distribution of Dirac nodes~\cite{lucht2023}, the $\mathbb Z_2$ invariant remains nontrivial and can therefore be generalized to any odd layer number. Although a Josephson spin current hasn't presently been experimentally demonstrated, spin current measurements are well-known experimentally. This work represents a natural step forward in the general realization of topological superconductivity using twists and interlayer (spin) currents that now allow for a non-trivial  $\mathbb{Z}$  or  $\mathbb{Z}_2$ index. Our results can further provide a framework  to continue this program of using stacking and twisting to realize higher order topological phases~\cite{benalcazar2017quantized,schindler2018higher}, which represents an interesting future direction.
With advancements in finite size flakes of layered singlet superconductors, their extension to triplet superconductors and combination with spin current effects offer a future outlook for twisted superconducting systems.

%%%%%%%%%%%%
%%%%%%%%%%%%
\appendix
%%%%%%%%%%%%
%%%%%%%%%%%%

\acknowledgements{We thank  Jennifer Cano, Marcel Franz, and Justin Wilson for useful discussions. This work is partially supported by NSF Career Grant No.~DMR- 1941569 and the Alfred P.~Sloan Foundation through a Sloan Research Fellowship (K.L., J.H.P.). Part of this work was performed in part at the Aspen Center for Physics, which is supported by the National Science Foundation Grant No.~PHY-2210452 (P.V., J.H.P.) as well as the Kavli Institute of Theoretical Physics that is supported in part by the National Science Foundation under Grants No.~NSF PHY-1748958 and PHY-2309135 (P.V., J.H.P.).}

\section{Projection of Two Layer Perturbations}
\label{appendix:A}
For two layers, we can write perturbations in the form $H_{\mathrm{pert}}^{\alpha\beta}(\bt{k}) = \sigma_\alpha\tau_\beta(\bt{h}_\beta(\bt{k})\cdot \bt{s} )$ where $\sigma_i$ represents the layer basis and $h$ is a parameter stemming, for example, from an external field. These perturbations are then applied to the two layer Hamiltonian in Eq.~\eqref{eq:bi_Ham0},

\begin{equation}
    H_2(\bt{k}) = \xi \tau_3 + \delta d_1(\bt{K}_N) \tau_1 s_3  -\alpha t d_1(\bt{K}_N) \tau_1 \sigma_3 s_3 + t \tau_3 \sigma_1. 
\end{equation}
From the zero energy basis (represented by basis vectors $\eta_i$ and $\zeta_i$) of the two layer Hamiltonian, we can take perturbations which satisfy $\mathcal{C}$ and $\mathcal{T}$ and find their non-zero contribution via perturbation theory. For the case where $h(\bt{k}) = h(-\bt{k})$, these projections are summarized in Table \ref{tab:even}. 

\begin{table}[h!]
\centering
\begin{tabular}{ c | c c c }
            & $\tau_3 s_0$                        &$\tau_3s_1$ & $\tau_2 s_2$                        \\ \hline 
   $\sigma_0$ & $-\sqrt{1-\alpha^2}h\zeta_3$ & $\sqrt{1-\alpha^2}h\eta_1$ & ---                      \\  
   $\sigma_1$ & $h\zeta_3\eta_3$             & $-h\eta_1$                 & ---                      \\
   $\sigma_2$ & $h\alpha \zeta_1$            & $-h\alpha \eta_2\zeta_1$   & $-h\alpha \zeta_2\eta_2$ \\
   $\sigma_3$ & $\alpha h\zeta_1\eta_3$      & $2\alpha h\zeta_3\eta_3$   & ---                      \\  
\end{tabular}
\caption{Projections to the zero energy basis for allows terms of the form $\sigma_i\tau_js_k$ with $\sigma_i$ along the vertical axis and $\tau_js_k$ along the horizontal. Here we assume that the perturbation is even under inversion.}
\label{tab:even}
\end{table}

We can also consider the case when the perturbation is odd under inversion where $h(\bt{k}) = - h(-\bt{k})$. In this instance, several more terms are permitted which are presented for $\sigma_i$ for $i \neq 2$ in Table \ref{tab:sigma_neq2}  and $\sigma_2$ in Table \ref{tab:sigma_2}. Terms denoted by $(*)$ have no contribution to first order, so the presented terms are second order perturbations, and $\tilde{h} = h \sqrt{1-\alpha^2}$

\begin{table}[h!]
\centering
\begin{tabular}{ c | c c c c c c}
        & $\tau_0s_1$                        & $\tau_0s_3$   & $\tau_1s_1$       & $\tau_1s_3$ & $\tau_2s_0$ & $\tau_3s_3$  \\ \hline
   $\sigma_0$ &  $\tilde{h}\eta_2\zeta_2 $  &  $-h\eta_3$ & $-h \eta_1\zeta_3$  & $\tilde{h}\zeta_1 $ & $h \zeta_2$  & $-h\eta_1\zeta_2 $ \\  
   $\sigma_1$ &  $-h\eta_2\zeta_2$  & $\tilde{h}\eta_3 $  & $\tilde{h}\eta_1\zeta_3 $  & $-h\zeta_1$  & $-\tilde{h}\zeta_2 $  & $\tilde{h}\eta_1\zeta_2 $ \\
   $\sigma_3$ &  $h\alpha \eta_1$  &  $\frac{h^2}{2t}\eta_3\zeta_3^*$ & $-\frac{h^2}{2t}\eta_3\zeta_3^*$  & $h\alpha\eta_3\zeta_3$ & $-\frac{h^2}{2t}\eta_3\zeta_3^*$  & $\frac{h^2}{2t}\eta_3\zeta_3^*$ 
\end{tabular}
\caption{Projections to the zero energy basis for allows terms of the form $\sigma_i\tau_js_k$ with $\sigma_i$ for $i=0,1,3$ along the vertical axis and $\tau_js_k$ along the horizontal. Here we assume that the perturbation is odd under inversion.}
\label{tab:sigma_neq2}
\end{table}

\begin{table}[h!]
\centering
\begin{tabular}{ c |c c }
            & $\tau_2s_2$                            &$\tau_3s_0$                     \\ \hline 
   $\sigma_2$ & $-\frac{h^2}{2t}\eta_3\zeta_3^*$                             & $\frac{h^2}{2t}\eta_3\zeta_3^*$       \\
\end{tabular}
\caption{Projections to the zero energy basis for allows terms of the form $\sigma_i\tau_js_k$ with $\sigma_i$ for $i=2$ along the vertical axis and $\tau_js_k$ along the horizontal for perturbations that are odd under inversion.}
\label{tab:sigma_2}
\end{table}

Let's highlight terms which may induce a Dirac mass. From the even terms, only $\sigma_2 \tau_1 s_1$ may be considered. Examined numerically, for a sufficiently large value of $h$, a gap can open in the spectrum for $\alpha < \alpha_{MA}$ (including zero) where the magic angle is shifted with the perturbation strength. After the magic angle, a nodal ring forms which then separates into two nodal rings which move away from one another along the perpendicular axis. The odd terms offer several possibilities which include those in the upper right of Table \ref{tab:sigma_neq2}:

\begin{itemize}
    \item $\sigma_0 \tau_2 s_0$, $\sigma_0 \tau_3 s_2$, $\sigma_1 \tau_2 s_0$, $\sigma_1 \tau_3 s_2$.
\end{itemize}

From left to right, the first term $\sigma_0 \tau_2 s_0$ corresponds to an equal spin current which opens a gap for all values of $\alpha$. As analysis of this term shows, it can indeed be treated as a Dirac mass which produces a topological phase characterized by a spin Chern number. The second term $\sigma_0 \tau_3 s_2$, however, forms nodal rings which then move towards and away from one another just as the nodal points in the unperturbed Hamiltonian. For the third term $\sigma_1 \tau_2 s_0$, a gap opens for $\alpha < \alpha_{MA}$ where the magic angle is shifted proportionally to the strength of the perturbation. Once the magic angle is reached, two Dirac nodes emerge and move along the perpendicular. The last term acts like the second term where nodal rings form for $\alpha < \alpha_{MA}$. Except once the magic angle is reached, nodes form instead of rings with move along the perpendicular as the twist angle increases.

\section{Singlet Superconductors}
\label{appendix:singlet}
Repeating the equal spin current analysis for a singlet superconductor likewise induces a spin phase. We'll start by rewriting Eq.~\eqref{eq:H_general} for a two layer system to form the Block Hamiltonian $H_{2,s}(\bt{k})$, 

\begin{eqnarray}
    & H_{2,s}(\bt{k}) = \xi \tau_3 + \delta \tau_1   -\alpha t \sigma_3\tau_1 + t \tau_3 \sigma_1 ,
    \label{eq:sing_H0}
\end{eqnarray}
by absorbing the spin component into the BW spinor $\Psi_{\bt{k}} \rightarrow (i s_2) \Psi_{\bt{k}}$. Applying the spin current transforms the Hamiltonian into $H_{2,s}(\bt{k}) \rightarrow \tilde{H}_{2,s}(\bt{k}) + H_{2,s,SJC}(\bt{k})$

\begin{eqnarray}
    & \tilde{H}_{2,s}(\bt{k}) = \xi \tau_3 + \delta \cos\frac{\varphi_s}{2} \tau_1  \nonumber \\
    & -\alpha t \cos\frac{\varphi_s}{2}\sigma_3\tau_1 + t \tau_3 \sigma_1 ,
    \label{eq:sing_Ham1}
\end{eqnarray}
\begin{eqnarray}
    & H_{2,s,SJC}(\bt{k}) =  -\delta \sin\frac{\varphi_s}{2}  \sigma_3\tau_2 s_3  + \alpha t \sin\frac{\varphi_s}{2} \tau_2 s_3 .
    \label{eq:sing_Ham2}
\end{eqnarray}
The imaginary component will break $SU(2)$ symmetry but have a remaining quantization axis in spin space. As a result, this mechanism will transform the classification from CI to AIII but not produce a topological index. Another important remark is that the order parameter must also transform into a triplet in order to satisfy Fermi-Dirac statistics, so simply inducing a spin component to a singlet order parameter is insufficient. Therefore, in order to transform the classification for a singlet superconductor to DIII requires additional contributions. For example, for a d-wave superconductor, the order parameter must transform to $d \rightarrow d + ip$ (or another triplet order parameter), and lose the spin quantization with an additional Rashba spin-orbit coupling for example.

\section{Josephson spin current in triplet superconductors}
\label{appendix:josephson}

Here we demonstrate that for c-axis interface, spin current is related to a spin phase of the order parameter by a Josephson relation (for the planar case see Ref \cite{Asano2006}). 
Let us first define the charge ($I_c$) and spin ($I_s$) currents as
\begin{equation}
\begin{gathered}
        I_c = \frac{ i e t}{\hbar} \sum_{\bf k,\sigma} [c^\dagger_{1,\sigma,\bf{k}}c_{2,\sigma,\bf{k}}-c^\dagger_{2,\sigma,\bf{k}}c_{1,\sigma,\bf{k}}] \\
        =
        \frac{e t}{\hbar} \sum_{\bf k,\sigma}
        \Psi_{\bf k}^\dagger 
        \sigma_2
        \Psi_{\bf k},
\\
        I_s^{\alpha=1,2,3} = \frac{t}{\hbar} \sum_{\bf k,\sigma}
        \Psi_{\bf k}^\dagger 
        \sigma_2 s_\alpha
        \Psi_{\bf k}.
\end{gathered}
\end{equation}
We will now compute the currents perturbatively in the interlayer tunneling. This approach is sufficient at low twist angles, since the main contribution comes away from nodes~\cite{Volkov2022_2}. In particular, we will consider here a $p_x$-wave order parameter $\hat{\Delta}\to \hat{\Delta} e^{i \varphi_s s_3 \sigma_3/2}$: $\Delta_0 \cos(\eta) \tau_1 s_3 \to \Delta_0 \cos(\eta) [ \cos(\varphi_s/2)\tau_1 s_3+\sin(\varphi_s/2)\sigma_3\tau_2]$, where $\eta$ is the angle on the Fermi surface. Note that such an operation is not possible for a singlet superconductor, since the spinful part of pairing has to be odd-parity. Consequently, application of a spinful phase difference to a singlet superconductor will result in a renormalization of $\hat{\Delta}$ by $\cos(\varphi_s/2)$ only, while the spinful term will vanish identically due to fermion anticommutation.

Continuing with the $z$-polarized spin current, it is then given by:

\begin{eqnarray}
\begin{gathered}
        I_{s}^3 = -\frac{ i t^2}{2\hbar} 
        T
        \sum_{\varepsilon_n,{\bf k}}
        {\rm Tr}
        \left[
        \sigma_2 s_3
        \hat{G}(i\varepsilon_n, {\bf k})
        \sigma_3 \tau_1
        \hat{G}(i\varepsilon_n, {\bf k})
        \right]
        \\
        =
         \frac{4 t^2 \Delta_0^2 \sin(\varphi_s)}{\hbar} 
         \int \frac{d\varepsilon d{\bf k}}{(2\pi)^3}
         \frac{\cos^2 \eta}{(\varepsilon^2+\xi^2+\Delta_0^2 \cos^2 \eta)^2}\Bigg\rvert_{T=0}
         \\
       = \frac{2t^2 \nu_0 \sin(\varphi_s)}{\hbar},
\end{gathered}
\end{eqnarray}
where $\nu_0$ is the density of states and the factor of $1/2$ is to compensate for the summation over all the momentum space (rather than half due to extended spinor structure). Note that the final result is only valid at $T\ll \Delta_0$. In the case of $I_{s}^3 \neq 0$, the contribution from equal and opposite spin current will therefore results in no charge current.

Importantly, one can further demonstrate that all other components of spin current and charge current vanish. Therefore, we have demonstrated that driving a spin current through a triplet twisted superconductor is equivalent to an application of spinful phase difference.

%\printbibliography
\bibliography{manuscript}

\begin{thebibliography}{46}%
\makeatletter
\providecommand \@ifxundefined [1]{%
 \@ifx{#1\undefined}
}%
\providecommand \@ifnum [1]{%
 \ifnum #1\expandafter \@firstoftwo
 \else \expandafter \@secondoftwo
 \fi
}%
\providecommand \@ifx [1]{%
 \ifx #1\expandafter \@firstoftwo
 \else \expandafter \@secondoftwo
 \fi
}%
\providecommand \natexlab [1]{#1}%
\providecommand \enquote  [1]{``#1''}%
\providecommand \bibnamefont  [1]{#1}%
\providecommand \bibfnamefont [1]{#1}%
\providecommand \citenamefont [1]{#1}%
\providecommand \href@noop [0]{\@secondoftwo}%
\providecommand \href [0]{\begingroup \@sanitize@url \@href}%
\providecommand \@href[1]{\@@startlink{#1}\@@href}%
\providecommand \@@href[1]{\endgroup#1\@@endlink}%
\providecommand \@sanitize@url [0]{\catcode `\\12\catcode `\$12\catcode
  `\&12\catcode `\#12\catcode `\^12\catcode `\_12\catcode `\%12\relax}%
\providecommand \@@startlink[1]{}%
\providecommand \@@endlink[0]{}%
\providecommand \url  [0]{\begingroup\@sanitize@url \@url }%
\providecommand \@url [1]{\endgroup\@href {#1}{\urlprefix }}%
\providecommand \urlprefix  [0]{URL }%
\providecommand \Eprint [0]{\href }%
\providecommand \doibase [0]{https://doi.org/}%
\providecommand \selectlanguage [0]{\@gobble}%
\providecommand \bibinfo  [0]{\@secondoftwo}%
\providecommand \bibfield  [0]{\@secondoftwo}%
\providecommand \translation [1]{[#1]}%
\providecommand \BibitemOpen [0]{}%
\providecommand \bibitemStop [0]{}%
\providecommand \bibitemNoStop [0]{.\EOS\space}%
\providecommand \EOS [0]{\spacefactor3000\relax}%
\providecommand \BibitemShut  [1]{\csname bibitem#1\endcsname}%
\let\auto@bib@innerbib\@empty
%</preamble>
\bibitem [{\citenamefont {Can}\ \emph {et~al.}(2021)\citenamefont {Can},
  \citenamefont {Tummuru}, \citenamefont {Day}, \citenamefont {Elfimov},
  \citenamefont {Damascelli},\ and\ \citenamefont {Franz}}]{Can2021}%
  \BibitemOpen
  \bibfield  {author} {\bibinfo {author} {\bibfnamefont {O.}~\bibnamefont
  {Can}}, \bibinfo {author} {\bibfnamefont {T.}~\bibnamefont {Tummuru}},
  \bibinfo {author} {\bibfnamefont {R.~P.}\ \bibnamefont {Day}}, \bibinfo
  {author} {\bibfnamefont {I.}~\bibnamefont {Elfimov}}, \bibinfo {author}
  {\bibfnamefont {A.}~\bibnamefont {Damascelli}},\ and\ \bibinfo {author}
  {\bibfnamefont {M.}~\bibnamefont {Franz}},\ }\bibfield  {title} {\bibinfo
  {title} {High-temperature topological superconductivity in twisted
  double-layer copper oxides},\ }\href
  {https://doi.org/10.1038/s41567-020-01142-7} {\bibfield  {journal} {\bibinfo
  {journal} {Nature Physics}\ }\textbf {\bibinfo {volume} {17}},\ \bibinfo
  {pages} {519} (\bibinfo {year} {2021})}\BibitemShut {NoStop}%
\bibitem [{\citenamefont {Volkov}\ \emph
  {et~al.}(2023{\natexlab{a}})\citenamefont {Volkov}, \citenamefont {Wilson},
  \citenamefont {Lucht},\ and\ \citenamefont {Pixley}}]{Volkov2022}%
  \BibitemOpen
  \bibfield  {author} {\bibinfo {author} {\bibfnamefont {P.~A.}\ \bibnamefont
  {Volkov}}, \bibinfo {author} {\bibfnamefont {J.~H.}\ \bibnamefont {Wilson}},
  \bibinfo {author} {\bibfnamefont {K.~P.}\ \bibnamefont {Lucht}},\ and\
  \bibinfo {author} {\bibfnamefont {J.~H.}\ \bibnamefont {Pixley}},\ }\bibfield
   {title} {\bibinfo {title} {Current- and field-induced topology in twisted
  nodal superconductors},\ }\href
  {https://doi.org/10.1103/PhysRevLett.130.186001} {\bibfield  {journal}
  {\bibinfo  {journal} {Phys. Rev. Lett.}\ }\textbf {\bibinfo {volume} {130}},\
  \bibinfo {pages} {186001} (\bibinfo {year} {2023}{\natexlab{a}})}\BibitemShut
  {NoStop}%
\bibitem [{\citenamefont {Margalit}\ \emph {et~al.}(2022)\citenamefont
  {Margalit}, \citenamefont {Yan}, \citenamefont {Franz},\ and\ \citenamefont
  {Oreg}}]{Margalit2022}%
  \BibitemOpen
  \bibfield  {author} {\bibinfo {author} {\bibfnamefont {G.}~\bibnamefont
  {Margalit}}, \bibinfo {author} {\bibfnamefont {B.}~\bibnamefont {Yan}},
  \bibinfo {author} {\bibfnamefont {M.}~\bibnamefont {Franz}},\ and\ \bibinfo
  {author} {\bibfnamefont {Y.}~\bibnamefont {Oreg}},\ }\bibfield  {title}
  {\bibinfo {title} {Chiral majorana modes via proximity to a twisted cuprate
  bilayer},\ }\href {https://doi.org/10.1103/PhysRevB.106.205424} {\bibfield
  {journal} {\bibinfo  {journal} {Phys. Rev. B}\ }\textbf {\bibinfo {volume}
  {106}},\ \bibinfo {pages} {205424} (\bibinfo {year} {2022})}\BibitemShut
  {NoStop}%
\bibitem [{\citenamefont {Yu}\ \emph {et~al.}(2019)\citenamefont {Yu},
  \citenamefont {Ma}, \citenamefont {Cai}, \citenamefont {Zhong}, \citenamefont
  {Ye}, \citenamefont {Shen}, \citenamefont {Gu}, \citenamefont {Chen},\ and\
  \citenamefont {Zhang}}]{Yu2019}%
  \BibitemOpen
  \bibfield  {author} {\bibinfo {author} {\bibfnamefont {Y.}~\bibnamefont
  {Yu}}, \bibinfo {author} {\bibfnamefont {L.}~\bibnamefont {Ma}}, \bibinfo
  {author} {\bibfnamefont {P.}~\bibnamefont {Cai}}, \bibinfo {author}
  {\bibfnamefont {R.}~\bibnamefont {Zhong}}, \bibinfo {author} {\bibfnamefont
  {C.}~\bibnamefont {Ye}}, \bibinfo {author} {\bibfnamefont {J.}~\bibnamefont
  {Shen}}, \bibinfo {author} {\bibfnamefont {G.~D.}\ \bibnamefont {Gu}},
  \bibinfo {author} {\bibfnamefont {X.~H.}\ \bibnamefont {Chen}},\ and\
  \bibinfo {author} {\bibfnamefont {Y.}~\bibnamefont {Zhang}},\ }\bibfield
  {title} {\bibinfo {title} {High-temperature superconductivity in monolayer
  bi2sr2cacu2o8+$\delta$},\ }\href {https://doi.org/10.1038/s41586-019-1718-x}
  {\bibfield  {journal} {\bibinfo  {journal} {Nature}\ }\textbf {\bibinfo
  {volume} {575}},\ \bibinfo {pages} {156} (\bibinfo {year}
  {2019})}\BibitemShut {NoStop}%
\bibitem [{\citenamefont {Volkov}\ \emph {et~al.}(2021)\citenamefont {Volkov},
  \citenamefont {Zhao}, \citenamefont {Poccia}, \citenamefont {Cui},
  \citenamefont {Kim},\ and\ \citenamefont {Pixley}}]{Volkov2021}%
  \BibitemOpen
  \bibfield  {author} {\bibinfo {author} {\bibfnamefont {P.~A.}\ \bibnamefont
  {Volkov}}, \bibinfo {author} {\bibfnamefont {S.~Y.~F.}\ \bibnamefont {Zhao}},
  \bibinfo {author} {\bibfnamefont {N.}~\bibnamefont {Poccia}}, \bibinfo
  {author} {\bibfnamefont {X.}~\bibnamefont {Cui}}, \bibinfo {author}
  {\bibfnamefont {P.}~\bibnamefont {Kim}},\ and\ \bibinfo {author}
  {\bibfnamefont {J.~H.}\ \bibnamefont {Pixley}},\ }\href
  {https://doi.org/10.48550/ARXIV.2108.13456} {\bibinfo {title} {Josephson
  effects in twisted nodal superconductors}} (\bibinfo {year}
  {2021})\BibitemShut {NoStop}%
\bibitem [{\citenamefont {Tummuru}\ \emph
  {et~al.}(2022{\natexlab{a}})\citenamefont {Tummuru}, \citenamefont {Plugge},\
  and\ \citenamefont {Franz}}]{Tummuru2022_2}%
  \BibitemOpen
  \bibfield  {author} {\bibinfo {author} {\bibfnamefont {T.}~\bibnamefont
  {Tummuru}}, \bibinfo {author} {\bibfnamefont {S.}~\bibnamefont {Plugge}},\
  and\ \bibinfo {author} {\bibfnamefont {M.}~\bibnamefont {Franz}},\ }\bibfield
   {title} {\bibinfo {title} {Josephson effects in twisted cuprate bilayers},\
  }\href {https://doi.org/10.1103/PhysRevB.105.064501} {\bibfield  {journal}
  {\bibinfo  {journal} {Phys. Rev. B}\ }\textbf {\bibinfo {volume} {105}},\
  \bibinfo {pages} {064501} (\bibinfo {year} {2022}{\natexlab{a}})}\BibitemShut
  {NoStop}%
\bibitem [{\citenamefont {Martini}\ \emph {et~al.}(2023)\citenamefont
  {Martini}, \citenamefont {Lee}, \citenamefont {Confalone}, \citenamefont
  {Shokri}, \citenamefont {Saggau}, \citenamefont {Wolf}, \citenamefont {Gu},
  \citenamefont {Watanabe}, \citenamefont {Taniguchi}, \citenamefont
  {Montemurro}, \citenamefont {Vinokur}, \citenamefont {Nielsch},\ and\
  \citenamefont {Poccia}}]{martini2023twisted}%
  \BibitemOpen
  \bibfield  {author} {\bibinfo {author} {\bibfnamefont {M.}~\bibnamefont
  {Martini}}, \bibinfo {author} {\bibfnamefont {Y.}~\bibnamefont {Lee}},
  \bibinfo {author} {\bibfnamefont {T.}~\bibnamefont {Confalone}}, \bibinfo
  {author} {\bibfnamefont {S.}~\bibnamefont {Shokri}}, \bibinfo {author}
  {\bibfnamefont {C.~N.}\ \bibnamefont {Saggau}}, \bibinfo {author}
  {\bibfnamefont {D.}~\bibnamefont {Wolf}}, \bibinfo {author} {\bibfnamefont
  {G.}~\bibnamefont {Gu}}, \bibinfo {author} {\bibfnamefont {K.}~\bibnamefont
  {Watanabe}}, \bibinfo {author} {\bibfnamefont {T.}~\bibnamefont {Taniguchi}},
  \bibinfo {author} {\bibfnamefont {D.}~\bibnamefont {Montemurro}}, \bibinfo
  {author} {\bibfnamefont {V.~M.}\ \bibnamefont {Vinokur}}, \bibinfo {author}
  {\bibfnamefont {K.}~\bibnamefont {Nielsch}},\ and\ \bibinfo {author}
  {\bibfnamefont {N.}~\bibnamefont {Poccia}},\ }\href@noop {} {\bibinfo {title}
  {Twisted cuprate van der waals heterostructures with controlled josephson
  coupling}} (\bibinfo {year} {2023}),\ \Eprint
  {https://arxiv.org/abs/2303.16029} {arXiv:2303.16029 [cond-mat.supr-con]}
  \BibitemShut {NoStop}%
\bibitem [{\citenamefont {Zhao}\ \emph {et~al.}(2023)\citenamefont {Zhao},
  \citenamefont {Cui}, \citenamefont {Volkov}, \citenamefont {Yoo},
  \citenamefont {Lee}, \citenamefont {Gardener}, \citenamefont {Akey},
  \citenamefont {Engelke}, \citenamefont {Ronen}, \citenamefont {Zhong} \emph
  {et~al.}}]{zhao2023time}%
  \BibitemOpen
  \bibfield  {author} {\bibinfo {author} {\bibfnamefont {S.~F.}\ \bibnamefont
  {Zhao}}, \bibinfo {author} {\bibfnamefont {X.}~\bibnamefont {Cui}}, \bibinfo
  {author} {\bibfnamefont {P.~A.}\ \bibnamefont {Volkov}}, \bibinfo {author}
  {\bibfnamefont {H.}~\bibnamefont {Yoo}}, \bibinfo {author} {\bibfnamefont
  {S.}~\bibnamefont {Lee}}, \bibinfo {author} {\bibfnamefont {J.~A.}\
  \bibnamefont {Gardener}}, \bibinfo {author} {\bibfnamefont {A.~J.}\
  \bibnamefont {Akey}}, \bibinfo {author} {\bibfnamefont {R.}~\bibnamefont
  {Engelke}}, \bibinfo {author} {\bibfnamefont {Y.}~\bibnamefont {Ronen}},
  \bibinfo {author} {\bibfnamefont {R.}~\bibnamefont {Zhong}}, \emph {et~al.},\
  }\bibfield  {title} {\bibinfo {title} {Time-reversal symmetry breaking
  superconductivity between twisted cuprate superconductors},\ }\href@noop {}
  {\bibfield  {journal} {\bibinfo  {journal} {Science}\ ,\ \bibinfo {pages}
  {eabl8371}} (\bibinfo {year} {2023})}\BibitemShut {NoStop}%
\bibitem [{\citenamefont {Tummuru}\ \emph {et~al.}(2021)\citenamefont
  {Tummuru}, \citenamefont {Can},\ and\ \citenamefont
  {Franz}}]{tummuru2020chiral}%
  \BibitemOpen
  \bibfield  {author} {\bibinfo {author} {\bibfnamefont {T.}~\bibnamefont
  {Tummuru}}, \bibinfo {author} {\bibfnamefont {O.}~\bibnamefont {Can}},\ and\
  \bibinfo {author} {\bibfnamefont {M.}~\bibnamefont {Franz}},\ }\bibfield
  {title} {\bibinfo {title} {Chiral $p$-wave superconductivity in a twisted
  array of proximitized quantum wires},\ }\href
  {https://doi.org/10.1103/PhysRevB.103.L100501} {\bibfield  {journal}
  {\bibinfo  {journal} {Phys. Rev. B}\ }\textbf {\bibinfo {volume} {103}},\
  \bibinfo {pages} {L100501} (\bibinfo {year} {2021})}\BibitemShut {NoStop}%
\bibitem [{\citenamefont {Volkov}\ \emph
  {et~al.}(2023{\natexlab{b}})\citenamefont {Volkov}, \citenamefont {Wilson},
  \citenamefont {Lucht},\ and\ \citenamefont {Pixley}}]{Volkov2022_2}%
  \BibitemOpen
  \bibfield  {author} {\bibinfo {author} {\bibfnamefont {P.~A.}\ \bibnamefont
  {Volkov}}, \bibinfo {author} {\bibfnamefont {J.~H.}\ \bibnamefont {Wilson}},
  \bibinfo {author} {\bibfnamefont {K.~P.}\ \bibnamefont {Lucht}},\ and\
  \bibinfo {author} {\bibfnamefont {J.~H.}\ \bibnamefont {Pixley}},\ }\bibfield
   {title} {\bibinfo {title} {Magic angles and correlations in twisted nodal
  superconductors},\ }\href {https://doi.org/10.1103/PhysRevB.107.174506}
  {\bibfield  {journal} {\bibinfo  {journal} {Phys. Rev. B}\ }\textbf {\bibinfo
  {volume} {107}},\ \bibinfo {pages} {174506} (\bibinfo {year}
  {2023}{\natexlab{b}})}\BibitemShut {NoStop}%
\bibitem [{\citenamefont {Liu}\ \emph {et~al.}(2023)\citenamefont {Liu},
  \citenamefont {Zhou}, \citenamefont {Zhang}, \citenamefont {Chen},\ and\
  \citenamefont {Yang}}]{liu2023}%
  \BibitemOpen
  \bibfield  {author} {\bibinfo {author} {\bibfnamefont {Y.-B.}\ \bibnamefont
  {Liu}}, \bibinfo {author} {\bibfnamefont {J.}~\bibnamefont {Zhou}}, \bibinfo
  {author} {\bibfnamefont {Y.}~\bibnamefont {Zhang}}, \bibinfo {author}
  {\bibfnamefont {W.-Q.}\ \bibnamefont {Chen}},\ and\ \bibinfo {author}
  {\bibfnamefont {F.}~\bibnamefont {Yang}},\ }\bibfield  {title} {\bibinfo
  {title} {Making chiral topological superconductors from nontopological
  superconductors through large angle twists},\ }\href
  {https://doi.org/10.1103/PhysRevB.108.064508} {\bibfield  {journal} {\bibinfo
   {journal} {Phys. Rev. B}\ }\textbf {\bibinfo {volume} {108}},\ \bibinfo
  {pages} {064508} (\bibinfo {year} {2023})}\BibitemShut {NoStop}%
\bibitem [{\citenamefont {Li}\ and\ \citenamefont {Liu}(2023)}]{Li2023}%
  \BibitemOpen
  \bibfield  {author} {\bibinfo {author} {\bibfnamefont {Y.-X.}\ \bibnamefont
  {Li}}\ and\ \bibinfo {author} {\bibfnamefont {C.-C.}\ \bibnamefont {Liu}},\
  }\bibfield  {title} {\bibinfo {title} {High-temperature majorana corner modes
  in a $d+i{d}^{\ensuremath{'}}$ superconductor heterostructure: Application to
  twisted bilayer cuprate superconductors},\ }\href
  {https://doi.org/10.1103/PhysRevB.107.235125} {\bibfield  {journal} {\bibinfo
   {journal} {Phys. Rev. B}\ }\textbf {\bibinfo {volume} {107}},\ \bibinfo
  {pages} {235125} (\bibinfo {year} {2023})}\BibitemShut {NoStop}%
\bibitem [{\citenamefont {Lin}\ \emph {et~al.}(2023)\citenamefont {Lin},
  \citenamefont {Huang},\ and\ \citenamefont {Lu}}]{Lin2023}%
  \BibitemOpen
  \bibfield  {author} {\bibinfo {author} {\bibfnamefont {C.}~\bibnamefont
  {Lin}}, \bibinfo {author} {\bibfnamefont {C.}~\bibnamefont {Huang}},\ and\
  \bibinfo {author} {\bibfnamefont {X.}~\bibnamefont {Lu}},\ }\bibfield
  {title} {\bibinfo {title} {Customizing topological phases in the twisted
  bilayer superconductors with even-parity pairings},\ }\href
  {http://iopscience.iop.org/article/10.1088/1674-1056/acd3e3} {\bibfield
  {journal} {\bibinfo  {journal} {Chinese Physics B}\ } (\bibinfo {year}
  {2023})}\BibitemShut {NoStop}%
\bibitem [{\citenamefont {Altland}\ and\ \citenamefont
  {Zirnbauer}(1997)}]{Altland1997}%
  \BibitemOpen
  \bibfield  {author} {\bibinfo {author} {\bibfnamefont {A.}~\bibnamefont
  {Altland}}\ and\ \bibinfo {author} {\bibfnamefont {M.~R.}\ \bibnamefont
  {Zirnbauer}},\ }\bibfield  {title} {\bibinfo {title} {Nonstandard symmetry
  classes in mesoscopic normal-superconducting hybrid structures},\ }\href
  {https://doi.org/10.1103/PhysRevB.55.1142} {\bibfield  {journal} {\bibinfo
  {journal} {Phys. Rev. B}\ }\textbf {\bibinfo {volume} {55}},\ \bibinfo
  {pages} {1142} (\bibinfo {year} {1997})}\BibitemShut {NoStop}%
\bibitem [{\citenamefont {Schnyder}\ \emph {et~al.}(2008)\citenamefont
  {Schnyder}, \citenamefont {Ryu}, \citenamefont {Furusaki},\ and\
  \citenamefont {Ludwig}}]{Schnyder2008}%
  \BibitemOpen
  \bibfield  {author} {\bibinfo {author} {\bibfnamefont {A.~P.}\ \bibnamefont
  {Schnyder}}, \bibinfo {author} {\bibfnamefont {S.}~\bibnamefont {Ryu}},
  \bibinfo {author} {\bibfnamefont {A.}~\bibnamefont {Furusaki}},\ and\
  \bibinfo {author} {\bibfnamefont {A.~W.~W.}\ \bibnamefont {Ludwig}},\
  }\bibfield  {title} {\bibinfo {title} {Classification of topological
  insulators and superconductors in three spatial dimensions},\ }\href
  {https://doi.org/10.1103/PhysRevB.78.195125} {\bibfield  {journal} {\bibinfo
  {journal} {Phys. Rev. B}\ }\textbf {\bibinfo {volume} {78}},\ \bibinfo
  {pages} {195125} (\bibinfo {year} {2008})}\BibitemShut {NoStop}%
\bibitem [{\citenamefont {Qi}\ \emph {et~al.}(2009)\citenamefont {Qi},
  \citenamefont {Hughes}, \citenamefont {Raghu},\ and\ \citenamefont
  {Zhang}}]{Qi2009}%
  \BibitemOpen
  \bibfield  {author} {\bibinfo {author} {\bibfnamefont {X.-L.}\ \bibnamefont
  {Qi}}, \bibinfo {author} {\bibfnamefont {T.~L.}\ \bibnamefont {Hughes}},
  \bibinfo {author} {\bibfnamefont {S.}~\bibnamefont {Raghu}},\ and\ \bibinfo
  {author} {\bibfnamefont {S.-C.}\ \bibnamefont {Zhang}},\ }\bibfield  {title}
  {\bibinfo {title} {Time-reversal-invariant topological superconductors and
  superfluids in two and three dimensions},\ }\href
  {https://doi.org/10.1103/PhysRevLett.102.187001} {\bibfield  {journal}
  {\bibinfo  {journal} {Phys. Rev. Lett.}\ }\textbf {\bibinfo {volume} {102}},\
  \bibinfo {pages} {187001} (\bibinfo {year} {2009})}\BibitemShut {NoStop}%
\bibitem [{\citenamefont {Ryu}\ \emph {et~al.}(2010)\citenamefont {Ryu},
  \citenamefont {Schnyder}, \citenamefont {Furusaki},\ and\ \citenamefont
  {Ludwig}}]{Ryu2010}%
  \BibitemOpen
  \bibfield  {author} {\bibinfo {author} {\bibfnamefont {S.}~\bibnamefont
  {Ryu}}, \bibinfo {author} {\bibfnamefont {A.~P.}\ \bibnamefont {Schnyder}},
  \bibinfo {author} {\bibfnamefont {A.}~\bibnamefont {Furusaki}},\ and\
  \bibinfo {author} {\bibfnamefont {A.~W.~W.}\ \bibnamefont {Ludwig}},\
  }\bibfield  {title} {\bibinfo {title} {Topological insulators and
  superconductors: tenfold way and dimensional hierarchy},\ }\href
  {https://doi.org/10.1088/1367-2630/12/6/065010} {\bibfield  {journal}
  {\bibinfo  {journal} {New Journal of Physics}\ }\textbf {\bibinfo {volume}
  {12}},\ \bibinfo {pages} {065010} (\bibinfo {year} {2010})}\BibitemShut
  {NoStop}%
\bibitem [{\citenamefont {Qi}\ \emph {et~al.}(2010)\citenamefont {Qi},
  \citenamefont {Hughes},\ and\ \citenamefont {Zhang}}]{Qi2010}%
  \BibitemOpen
  \bibfield  {author} {\bibinfo {author} {\bibfnamefont {X.-L.}\ \bibnamefont
  {Qi}}, \bibinfo {author} {\bibfnamefont {T.~L.}\ \bibnamefont {Hughes}},\
  and\ \bibinfo {author} {\bibfnamefont {S.-C.}\ \bibnamefont {Zhang}},\
  }\bibfield  {title} {\bibinfo {title} {Topological invariants for the fermi
  surface of a time-reversal-invariant superconductor},\ }\href
  {https://doi.org/10.1103/PhysRevB.81.134508} {\bibfield  {journal} {\bibinfo
  {journal} {Phys. Rev. B}\ }\textbf {\bibinfo {volume} {81}},\ \bibinfo
  {pages} {134508} (\bibinfo {year} {2010})}\BibitemShut {NoStop}%
\bibitem [{\citenamefont {Deng}\ \emph {et~al.}(2012)\citenamefont {Deng},
  \citenamefont {Viola},\ and\ \citenamefont {Ortiz}}]{Deng2012}%
  \BibitemOpen
  \bibfield  {author} {\bibinfo {author} {\bibfnamefont {S.}~\bibnamefont
  {Deng}}, \bibinfo {author} {\bibfnamefont {L.}~\bibnamefont {Viola}},\ and\
  \bibinfo {author} {\bibfnamefont {G.}~\bibnamefont {Ortiz}},\ }\bibfield
  {title} {\bibinfo {title} {Majorana modes in time-reversal invariant $s$-wave
  topological superconductors},\ }\href
  {https://doi.org/10.1103/PhysRevLett.108.036803} {\bibfield  {journal}
  {\bibinfo  {journal} {Phys. Rev. Lett.}\ }\textbf {\bibinfo {volume} {108}},\
  \bibinfo {pages} {036803} (\bibinfo {year} {2012})}\BibitemShut {NoStop}%
\bibitem [{\citenamefont {Zhang}\ \emph {et~al.}(2013)\citenamefont {Zhang},
  \citenamefont {Kane},\ and\ \citenamefont {Mele}}]{Zhang2013}%
  \BibitemOpen
  \bibfield  {author} {\bibinfo {author} {\bibfnamefont {F.}~\bibnamefont
  {Zhang}}, \bibinfo {author} {\bibfnamefont {C.~L.}\ \bibnamefont {Kane}},\
  and\ \bibinfo {author} {\bibfnamefont {E.~J.}\ \bibnamefont {Mele}},\
  }\bibfield  {title} {\bibinfo {title} {Time-reversal-invariant topological
  superconductivity and majorana kramers pairs},\ }\href
  {https://doi.org/10.1103/PhysRevLett.111.056402} {\bibfield  {journal}
  {\bibinfo  {journal} {Phys. Rev. Lett.}\ }\textbf {\bibinfo {volume} {111}},\
  \bibinfo {pages} {056402} (\bibinfo {year} {2013})}\BibitemShut {NoStop}%
\bibitem [{\citenamefont {Qin}\ \emph {et~al.}(2022)\citenamefont {Qin},
  \citenamefont {Fang}, \citenamefont {Zhang},\ and\ \citenamefont
  {Hu}}]{Qin2022}%
  \BibitemOpen
  \bibfield  {author} {\bibinfo {author} {\bibfnamefont {S.}~\bibnamefont
  {Qin}}, \bibinfo {author} {\bibfnamefont {C.}~\bibnamefont {Fang}}, \bibinfo
  {author} {\bibfnamefont {F.-C.}\ \bibnamefont {Zhang}},\ and\ \bibinfo
  {author} {\bibfnamefont {J.}~\bibnamefont {Hu}},\ }\bibfield  {title}
  {\bibinfo {title} {Topological superconductivity in an extended $s$-wave
  superconductor and its implication to iron-based superconductors},\ }\href
  {https://doi.org/10.1103/PhysRevX.12.011030} {\bibfield  {journal} {\bibinfo
  {journal} {Phys. Rev. X}\ }\textbf {\bibinfo {volume} {12}},\ \bibinfo
  {pages} {011030} (\bibinfo {year} {2022})}\BibitemShut {NoStop}%
\bibitem [{\citenamefont {Zhang}\ and\ \citenamefont
  {Das~Sarma}(2021)}]{Zhang2021}%
  \BibitemOpen
  \bibfield  {author} {\bibinfo {author} {\bibfnamefont {R.-X.}\ \bibnamefont
  {Zhang}}\ and\ \bibinfo {author} {\bibfnamefont {S.}~\bibnamefont
  {Das~Sarma}},\ }\bibfield  {title} {\bibinfo {title} {Intrinsic
  time-reversal-invariant topological superconductivity in thin films of
  iron-based superconductors},\ }\href
  {https://doi.org/10.1103/PhysRevLett.126.137001} {\bibfield  {journal}
  {\bibinfo  {journal} {Phys. Rev. Lett.}\ }\textbf {\bibinfo {volume} {126}},\
  \bibinfo {pages} {137001} (\bibinfo {year} {2021})}\BibitemShut {NoStop}%
\bibitem [{\citenamefont {Brydon}\ \emph {et~al.}(2008)\citenamefont {Brydon},
  \citenamefont {Kastening}, \citenamefont {Morr},\ and\ \citenamefont
  {Manske}}]{Brydon2008}%
  \BibitemOpen
  \bibfield  {author} {\bibinfo {author} {\bibfnamefont {P.~M.~R.}\
  \bibnamefont {Brydon}}, \bibinfo {author} {\bibfnamefont {B.}~\bibnamefont
  {Kastening}}, \bibinfo {author} {\bibfnamefont {D.~K.}\ \bibnamefont
  {Morr}},\ and\ \bibinfo {author} {\bibfnamefont {D.}~\bibnamefont {Manske}},\
  }\bibfield  {title} {\bibinfo {title} {Interplay of ferromagnetism and
  triplet superconductivity in a josephson junction},\ }\href
  {https://doi.org/10.1103/PhysRevB.77.104504} {\bibfield  {journal} {\bibinfo
  {journal} {Phys. Rev. B}\ }\textbf {\bibinfo {volume} {77}},\ \bibinfo
  {pages} {104504} (\bibinfo {year} {2008})}\BibitemShut {NoStop}%
\bibitem [{\citenamefont {Alidoust}\ \emph {et~al.}(2010)\citenamefont
  {Alidoust}, \citenamefont {Linder}, \citenamefont {Rashedi}, \citenamefont
  {Yokoyama},\ and\ \citenamefont {Sudb\o{}}}]{Alidoust2010}%
  \BibitemOpen
  \bibfield  {author} {\bibinfo {author} {\bibfnamefont {M.}~\bibnamefont
  {Alidoust}}, \bibinfo {author} {\bibfnamefont {J.}~\bibnamefont {Linder}},
  \bibinfo {author} {\bibfnamefont {G.}~\bibnamefont {Rashedi}}, \bibinfo
  {author} {\bibfnamefont {T.}~\bibnamefont {Yokoyama}},\ and\ \bibinfo
  {author} {\bibfnamefont {A.}~\bibnamefont {Sudb\o{}}},\ }\bibfield  {title}
  {\bibinfo {title} {Spin-polarized josephson current in
  superconductor/ferromagnet/superconductor junctions with inhomogeneous
  magnetization},\ }\href {https://doi.org/10.1103/PhysRevB.81.014512}
  {\bibfield  {journal} {\bibinfo  {journal} {Phys. Rev. B}\ }\textbf {\bibinfo
  {volume} {81}},\ \bibinfo {pages} {014512} (\bibinfo {year}
  {2010})}\BibitemShut {NoStop}%
\bibitem [{\citenamefont {Hikino}\ and\ \citenamefont
  {Yunoki}(2013)}]{Hikino2013}%
  \BibitemOpen
  \bibfield  {author} {\bibinfo {author} {\bibfnamefont {S.}~\bibnamefont
  {Hikino}}\ and\ \bibinfo {author} {\bibfnamefont {S.}~\bibnamefont
  {Yunoki}},\ }\bibfield  {title} {\bibinfo {title} {Long-range spin current
  driven by superconducting phase difference in a josephson junction with
  double layer ferromagnets},\ }\href
  {https://doi.org/10.1103/PhysRevLett.110.237003} {\bibfield  {journal}
  {\bibinfo  {journal} {Phys. Rev. Lett.}\ }\textbf {\bibinfo {volume} {110}},\
  \bibinfo {pages} {237003} (\bibinfo {year} {2013})}\BibitemShut {NoStop}%
\bibitem [{\citenamefont {Costa}\ and\ \citenamefont
  {Fabian}(2020)}]{Costa2020}%
  \BibitemOpen
  \bibfield  {author} {\bibinfo {author} {\bibfnamefont {A.}~\bibnamefont
  {Costa}}\ and\ \bibinfo {author} {\bibfnamefont {J.}~\bibnamefont {Fabian}},\
  }\bibfield  {title} {\bibinfo {title} {Anomalous josephson hall effect charge
  and transverse spin currents in
  superconductor/ferromagnetic-insulator/superconductor junctions},\ }\href
  {https://doi.org/10.1103/PhysRevB.101.104508} {\bibfield  {journal} {\bibinfo
   {journal} {Phys. Rev. B}\ }\textbf {\bibinfo {volume} {101}},\ \bibinfo
  {pages} {104508} (\bibinfo {year} {2020})}\BibitemShut {NoStop}%
\bibitem [{\citenamefont {Dai}\ \emph {et~al.}(2022)\citenamefont {Dai},
  \citenamefont {Mao},\ and\ \citenamefont {Sun}}]{Dai2022}%
  \BibitemOpen
  \bibfield  {author} {\bibinfo {author} {\bibfnamefont {Y.-X.}\ \bibnamefont
  {Dai}}, \bibinfo {author} {\bibfnamefont {Y.}~\bibnamefont {Mao}},\ and\
  \bibinfo {author} {\bibfnamefont {Q.-F.}\ \bibnamefont {Sun}},\ }\bibfield
  {title} {\bibinfo {title} {Spin transport in a normal metal--ising
  superconductor junction},\ }\href
  {https://doi.org/10.1103/PhysRevB.106.184513} {\bibfield  {journal} {\bibinfo
   {journal} {Phys. Rev. B}\ }\textbf {\bibinfo {volume} {106}},\ \bibinfo
  {pages} {184513} (\bibinfo {year} {2022})}\BibitemShut {NoStop}%
\bibitem [{\citenamefont {Keizer}\ \emph {et~al.}(2006)\citenamefont {Keizer},
  \citenamefont {Goennenwein}, \citenamefont {Klapwijk}, \citenamefont {Miao},
  \citenamefont {Xiao},\ and\ \citenamefont {Gupta}}]{Keizer2006}%
  \BibitemOpen
  \bibfield  {author} {\bibinfo {author} {\bibfnamefont {R.~S.}\ \bibnamefont
  {Keizer}}, \bibinfo {author} {\bibfnamefont {S.~T.~B.}\ \bibnamefont
  {Goennenwein}}, \bibinfo {author} {\bibfnamefont {T.~M.}\ \bibnamefont
  {Klapwijk}}, \bibinfo {author} {\bibfnamefont {G.}~\bibnamefont {Miao}},
  \bibinfo {author} {\bibfnamefont {G.}~\bibnamefont {Xiao}},\ and\ \bibinfo
  {author} {\bibfnamefont {A.}~\bibnamefont {Gupta}},\ }\bibfield  {title}
  {\bibinfo {title} {A spin triplet supercurrent through the half-metallic
  ferromagnet cro2},\ }\href {https://doi.org/10.1038/nature04499} {\bibfield
  {journal} {\bibinfo  {journal} {Nature}\ }\textbf {\bibinfo {volume} {439}},\
  \bibinfo {pages} {825} (\bibinfo {year} {2006})}\BibitemShut {NoStop}%
\bibitem [{\citenamefont {Anwar}\ \emph {et~al.}(2010)\citenamefont {Anwar},
  \citenamefont {Czeschka}, \citenamefont {Hesselberth}, \citenamefont
  {Porcu},\ and\ \citenamefont {Aarts}}]{Anwar2010}%
  \BibitemOpen
  \bibfield  {author} {\bibinfo {author} {\bibfnamefont {M.~S.}\ \bibnamefont
  {Anwar}}, \bibinfo {author} {\bibfnamefont {F.}~\bibnamefont {Czeschka}},
  \bibinfo {author} {\bibfnamefont {M.}~\bibnamefont {Hesselberth}}, \bibinfo
  {author} {\bibfnamefont {M.}~\bibnamefont {Porcu}},\ and\ \bibinfo {author}
  {\bibfnamefont {J.}~\bibnamefont {Aarts}},\ }\bibfield  {title} {\bibinfo
  {title} {Long-range supercurrents through half-metallic ferromagnetic
  ${\text{cro}}_{2}$},\ }\href {https://doi.org/10.1103/PhysRevB.82.100501}
  {\bibfield  {journal} {\bibinfo  {journal} {Phys. Rev. B}\ }\textbf {\bibinfo
  {volume} {82}},\ \bibinfo {pages} {100501} (\bibinfo {year}
  {2010})}\BibitemShut {NoStop}%
\bibitem [{\citenamefont {Singh}\ \emph {et~al.}(2015)\citenamefont {Singh},
  \citenamefont {Voltan}, \citenamefont {Lahabi},\ and\ \citenamefont
  {Aarts}}]{Singh2015}%
  \BibitemOpen
  \bibfield  {author} {\bibinfo {author} {\bibfnamefont {A.}~\bibnamefont
  {Singh}}, \bibinfo {author} {\bibfnamefont {S.}~\bibnamefont {Voltan}},
  \bibinfo {author} {\bibfnamefont {K.}~\bibnamefont {Lahabi}},\ and\ \bibinfo
  {author} {\bibfnamefont {J.}~\bibnamefont {Aarts}},\ }\bibfield  {title}
  {\bibinfo {title} {Colossal proximity effect in a superconducting triplet
  spin valve based on the half-metallic ferromagnet ${\mathrm{cro}}_{2}$},\
  }\href {https://doi.org/10.1103/PhysRevX.5.021019} {\bibfield  {journal}
  {\bibinfo  {journal} {Phys. Rev. X}\ }\textbf {\bibinfo {volume} {5}},\
  \bibinfo {pages} {021019} (\bibinfo {year} {2015})}\BibitemShut {NoStop}%
\bibitem [{\citenamefont {Han}\ \emph {et~al.}(2020)\citenamefont {Han},
  \citenamefont {Maekawa},\ and\ \citenamefont {Xie}}]{Han2020Review}%
  \BibitemOpen
  \bibfield  {author} {\bibinfo {author} {\bibfnamefont {W.}~\bibnamefont
  {Han}}, \bibinfo {author} {\bibfnamefont {S.}~\bibnamefont {Maekawa}},\ and\
  \bibinfo {author} {\bibfnamefont {X.-C.}\ \bibnamefont {Xie}},\ }\bibfield
  {title} {\bibinfo {title} {Spin current as a probe of quantum materials},\
  }\href {https://doi.org/10.1038/s41563-019-0456-7} {\bibfield  {journal}
  {\bibinfo  {journal} {Nature Materials}\ }\textbf {\bibinfo {volume} {19}},\
  \bibinfo {pages} {139} (\bibinfo {year} {2020})}\BibitemShut {NoStop}%
\bibitem [{\citenamefont {Sanchez-Manzano}\ \emph {et~al.}(2022)\citenamefont
  {Sanchez-Manzano}, \citenamefont {Mesoraca}, \citenamefont {Cuellar},
  \citenamefont {Cabero}, \citenamefont {Rouco}, \citenamefont {Orfila},
  \citenamefont {Palermo}, \citenamefont {Balan}, \citenamefont {Marcano},
  \citenamefont {Sander}, \citenamefont {Rocci}, \citenamefont
  {Garcia-Barriocanal}, \citenamefont {Gallego}, \citenamefont {Tornos},
  \citenamefont {Rivera}, \citenamefont {Mompean}, \citenamefont
  {Garcia-Hernandez}, \citenamefont {Gonzalez-Calbet}, \citenamefont {Leon},
  \citenamefont {Valencia}, \citenamefont {Feuillet-Palma}, \citenamefont
  {Bergeal}, \citenamefont {Buzdin}, \citenamefont {Lesueur}, \citenamefont
  {Villegas},\ and\ \citenamefont {Santamaria}}]{Sanchez-Manzano2022}%
  \BibitemOpen
  \bibfield  {author} {\bibinfo {author} {\bibfnamefont {D.}~\bibnamefont
  {Sanchez-Manzano}}, \bibinfo {author} {\bibfnamefont {S.}~\bibnamefont
  {Mesoraca}}, \bibinfo {author} {\bibfnamefont {F.~A.}\ \bibnamefont
  {Cuellar}}, \bibinfo {author} {\bibfnamefont {M.}~\bibnamefont {Cabero}},
  \bibinfo {author} {\bibfnamefont {V.}~\bibnamefont {Rouco}}, \bibinfo
  {author} {\bibfnamefont {G.}~\bibnamefont {Orfila}}, \bibinfo {author}
  {\bibfnamefont {X.}~\bibnamefont {Palermo}}, \bibinfo {author} {\bibfnamefont
  {A.}~\bibnamefont {Balan}}, \bibinfo {author} {\bibfnamefont
  {L.}~\bibnamefont {Marcano}}, \bibinfo {author} {\bibfnamefont
  {A.}~\bibnamefont {Sander}}, \bibinfo {author} {\bibfnamefont
  {M.}~\bibnamefont {Rocci}}, \bibinfo {author} {\bibfnamefont
  {J.}~\bibnamefont {Garcia-Barriocanal}}, \bibinfo {author} {\bibfnamefont
  {F.}~\bibnamefont {Gallego}}, \bibinfo {author} {\bibfnamefont
  {J.}~\bibnamefont {Tornos}}, \bibinfo {author} {\bibfnamefont
  {A.}~\bibnamefont {Rivera}}, \bibinfo {author} {\bibfnamefont
  {F.}~\bibnamefont {Mompean}}, \bibinfo {author} {\bibfnamefont
  {M.}~\bibnamefont {Garcia-Hernandez}}, \bibinfo {author} {\bibfnamefont
  {J.~M.}\ \bibnamefont {Gonzalez-Calbet}}, \bibinfo {author} {\bibfnamefont
  {C.}~\bibnamefont {Leon}}, \bibinfo {author} {\bibfnamefont {S.}~\bibnamefont
  {Valencia}}, \bibinfo {author} {\bibfnamefont {C.}~\bibnamefont
  {Feuillet-Palma}}, \bibinfo {author} {\bibfnamefont {N.}~\bibnamefont
  {Bergeal}}, \bibinfo {author} {\bibfnamefont {A.~I.}\ \bibnamefont {Buzdin}},
  \bibinfo {author} {\bibfnamefont {J.}~\bibnamefont {Lesueur}}, \bibinfo
  {author} {\bibfnamefont {J.~E.}\ \bibnamefont {Villegas}},\ and\ \bibinfo
  {author} {\bibfnamefont {J.}~\bibnamefont {Santamaria}},\ }\bibfield  {title}
  {\bibinfo {title} {Extremely long-range, high-temperature josephson coupling
  across a half-metallic ferromagnet},\ }\href
  {https://doi.org/10.1038/s41563-021-01162-5} {\bibfield  {journal} {\bibinfo
  {journal} {Nature Materials}\ }\textbf {\bibinfo {volume} {21}},\ \bibinfo
  {pages} {188} (\bibinfo {year} {2022})}\BibitemShut {NoStop}%
\bibitem [{\citenamefont {Kumawat}\ \emph {et~al.}(2023)\citenamefont
  {Kumawat}, \citenamefont {Dwivedi}, \citenamefont {Su}, \citenamefont {Shyu},
  \citenamefont {Chien}, \citenamefont {Su}, \citenamefont {Chung},
  \citenamefont {Fernandez}, \citenamefont {Sun}, \citenamefont {Hsu},
  \citenamefont {Yang},\ and\ \citenamefont {Chou}}]{Kumawat2023}%
  \BibitemOpen
  \bibfield  {author} {\bibinfo {author} {\bibfnamefont {S.~M.}\ \bibnamefont
  {Kumawat}}, \bibinfo {author} {\bibfnamefont {G.~D.}\ \bibnamefont
  {Dwivedi}}, \bibinfo {author} {\bibfnamefont {P.~F.}\ \bibnamefont {Su}},
  \bibinfo {author} {\bibfnamefont {W.~S.}\ \bibnamefont {Shyu}}, \bibinfo
  {author} {\bibfnamefont {Y.~H.}\ \bibnamefont {Chien}}, \bibinfo {author}
  {\bibfnamefont {P.~W.}\ \bibnamefont {Su}}, \bibinfo {author} {\bibfnamefont
  {C.~M.}\ \bibnamefont {Chung}}, \bibinfo {author} {\bibfnamefont {N.~D.~B.}\
  \bibnamefont {Fernandez}}, \bibinfo {author} {\bibfnamefont {S.~J.}\
  \bibnamefont {Sun}}, \bibinfo {author} {\bibfnamefont {C.-H.}\ \bibnamefont
  {Hsu}}, \bibinfo {author} {\bibfnamefont {S.}~\bibnamefont {Yang}},\ and\
  \bibinfo {author} {\bibfnamefont {H.}~\bibnamefont {Chou}},\ }\bibfield
  {title} {\bibinfo {title} {Magnetic field enhancement in critical current and
  possible triplet superconductivity in lsmo/ybco/lsmo heterostructures},\
  }\href {https://doi.org/10.1021/acs.jpcc.2c08620} {\bibfield  {journal}
  {\bibinfo  {journal} {The Journal of Physical Chemistry C}\ }\textbf
  {\bibinfo {volume} {127}},\ \bibinfo {pages} {6861} (\bibinfo {year}
  {2023})}\BibitemShut {NoStop}%
\bibitem [{\citenamefont {Hu}\ \emph {et~al.}(2023)\citenamefont {Hu},
  \citenamefont {Wang}, \citenamefont {Wang}, \citenamefont {Zhang},
  \citenamefont {Feng}, \citenamefont {Wang}, \citenamefont {Niu},
  \citenamefont {Zhang},\ and\ \citenamefont {Xiang}}]{Hu2023}%
  \BibitemOpen
  \bibfield  {author} {\bibinfo {author} {\bibfnamefont {G.}~\bibnamefont
  {Hu}}, \bibinfo {author} {\bibfnamefont {C.}~\bibnamefont {Wang}}, \bibinfo
  {author} {\bibfnamefont {S.}~\bibnamefont {Wang}}, \bibinfo {author}
  {\bibfnamefont {Y.}~\bibnamefont {Zhang}}, \bibinfo {author} {\bibfnamefont
  {Y.}~\bibnamefont {Feng}}, \bibinfo {author} {\bibfnamefont {Z.}~\bibnamefont
  {Wang}}, \bibinfo {author} {\bibfnamefont {Q.}~\bibnamefont {Niu}}, \bibinfo
  {author} {\bibfnamefont {Z.}~\bibnamefont {Zhang}},\ and\ \bibinfo {author}
  {\bibfnamefont {B.}~\bibnamefont {Xiang}},\ }\bibfield  {title} {\bibinfo
  {title} {Long-range skin josephson supercurrent across a van der waals
  ferromagnet},\ }\href {https://doi.org/10.1038/s41467-023-37603-9} {\bibfield
   {journal} {\bibinfo  {journal} {Nature Communications}\ }\textbf {\bibinfo
  {volume} {14}},\ \bibinfo {pages} {1779} (\bibinfo {year}
  {2023})}\BibitemShut {NoStop}%
\bibitem [{\citenamefont {Jeon}\ \emph {et~al.}(2018)\citenamefont {Jeon},
  \citenamefont {Ciccarelli}, \citenamefont {Ferguson}, \citenamefont
  {Kurebayashi}, \citenamefont {Cohen}, \citenamefont {Montiel}, \citenamefont
  {Eschrig}, \citenamefont {Robinson},\ and\ \citenamefont
  {Blamire}}]{Jeon2018}%
  \BibitemOpen
  \bibfield  {author} {\bibinfo {author} {\bibfnamefont {K.-R.}\ \bibnamefont
  {Jeon}}, \bibinfo {author} {\bibfnamefont {C.}~\bibnamefont {Ciccarelli}},
  \bibinfo {author} {\bibfnamefont {A.~J.}\ \bibnamefont {Ferguson}}, \bibinfo
  {author} {\bibfnamefont {H.}~\bibnamefont {Kurebayashi}}, \bibinfo {author}
  {\bibfnamefont {L.~F.}\ \bibnamefont {Cohen}}, \bibinfo {author}
  {\bibfnamefont {X.}~\bibnamefont {Montiel}}, \bibinfo {author} {\bibfnamefont
  {M.}~\bibnamefont {Eschrig}}, \bibinfo {author} {\bibfnamefont {J.~W.~A.}\
  \bibnamefont {Robinson}},\ and\ \bibinfo {author} {\bibfnamefont {M.~G.}\
  \bibnamefont {Blamire}},\ }\bibfield  {title} {\bibinfo {title} {Enhanced
  spin pumping into superconductors provides evidence for superconducting pure
  spin currents},\ }\href {https://doi.org/10.1038/s41563-018-0058-9}
  {\bibfield  {journal} {\bibinfo  {journal} {Nature Materials}\ }\textbf
  {\bibinfo {volume} {17}},\ \bibinfo {pages} {499} (\bibinfo {year}
  {2018})}\BibitemShut {NoStop}%
\bibitem [{\citenamefont {Asano}(2006)}]{Asano2006}%
  \BibitemOpen
  \bibfield  {author} {\bibinfo {author} {\bibfnamefont {Y.}~\bibnamefont
  {Asano}},\ }\bibfield  {title} {\bibinfo {title} {Josephson spin current in
  triplet superconductor junctions},\ }\href
  {https://doi.org/10.1103/PhysRevB.74.220501} {\bibfield  {journal} {\bibinfo
  {journal} {Phys. Rev. B}\ }\textbf {\bibinfo {volume} {74}},\ \bibinfo
  {pages} {220501} (\bibinfo {year} {2006})}\BibitemShut {NoStop}%
\bibitem [{\citenamefont {Rashedi}\ and\ \citenamefont
  {Kolesnichenko}(2007)}]{Rashedi2007}%
  \BibitemOpen
  \bibfield  {author} {\bibinfo {author} {\bibfnamefont {G.}~\bibnamefont
  {Rashedi}}\ and\ \bibinfo {author} {\bibfnamefont {Y.}~\bibnamefont
  {Kolesnichenko}},\ }\bibfield  {title} {\bibinfo {title} {Spin polarized
  transport in the weak link between f-wave superconductors},\ }\href
  {https://doi.org/https://doi.org/10.1016/j.physc.2006.10.001} {\bibfield
  {journal} {\bibinfo  {journal} {Physica C: Superconductivity}\ }\textbf
  {\bibinfo {volume} {451}},\ \bibinfo {pages} {31} (\bibinfo {year}
  {2007})}\BibitemShut {NoStop}%
\bibitem [{\citenamefont {Brydon}\ and\ \citenamefont
  {Manske}(2009)}]{Brydon2009}%
  \BibitemOpen
  \bibfield  {author} {\bibinfo {author} {\bibfnamefont {P.~M.~R.}\
  \bibnamefont {Brydon}}\ and\ \bibinfo {author} {\bibfnamefont
  {D.}~\bibnamefont {Manske}},\ }\bibfield  {title} {\bibinfo {title}
  {0-$\ensuremath{\pi}$ transition in magnetic triplet superconductor josephson
  junctions},\ }\href {https://doi.org/10.1103/PhysRevLett.103.147001}
  {\bibfield  {journal} {\bibinfo  {journal} {Phys. Rev. Lett.}\ }\textbf
  {\bibinfo {volume} {103}},\ \bibinfo {pages} {147001} (\bibinfo {year}
  {2009})}\BibitemShut {NoStop}%
\bibitem [{\citenamefont {Tummuru}\ \emph
  {et~al.}(2022{\natexlab{b}})\citenamefont {Tummuru}, \citenamefont
  {Lantagne-Hurtubise},\ and\ \citenamefont {Franz}}]{Tummuru2022}%
  \BibitemOpen
  \bibfield  {author} {\bibinfo {author} {\bibfnamefont {T.}~\bibnamefont
  {Tummuru}}, \bibinfo {author} {\bibfnamefont {E.}~\bibnamefont
  {Lantagne-Hurtubise}},\ and\ \bibinfo {author} {\bibfnamefont
  {M.}~\bibnamefont {Franz}},\ }\bibfield  {title} {\bibinfo {title} {Twisted
  multilayer nodal superconductors},\ }\href
  {https://doi.org/10.1103/PhysRevB.106.014520} {\bibfield  {journal} {\bibinfo
   {journal} {Phys. Rev. B}\ }\textbf {\bibinfo {volume} {106}},\ \bibinfo
  {pages} {014520} (\bibinfo {year} {2022}{\natexlab{b}})}\BibitemShut
  {NoStop}%
\bibitem [{\citenamefont {Lucht}\ \emph {et~al.}(2023)\citenamefont {Lucht},
  \citenamefont {Pixley},\ and\ \citenamefont {Volkov}}]{lucht2023}%
  \BibitemOpen
  \bibfield  {author} {\bibinfo {author} {\bibfnamefont {K.~P.}\ \bibnamefont
  {Lucht}}, \bibinfo {author} {\bibfnamefont {J.~H.}\ \bibnamefont {Pixley}},\
  and\ \bibinfo {author} {\bibfnamefont {P.~A.}\ \bibnamefont {Volkov}},\
  }\href@noop {} {\bibinfo {title} {Topological superconductivity in twisted
  flakes of nodal superconductors}} (\bibinfo {year} {2023}),\ \Eprint
  {https://arxiv.org/abs/2312.13367} {arXiv:2312.13367 [cond-mat.supr-con]}
  \BibitemShut {NoStop}%
\bibitem [{\citenamefont {Mao}\ and\ \citenamefont {Sun}(2022)}]{Mao2022}%
  \BibitemOpen
  \bibfield  {author} {\bibinfo {author} {\bibfnamefont {Y.}~\bibnamefont
  {Mao}}\ and\ \bibinfo {author} {\bibfnamefont {Q.-F.}\ \bibnamefont {Sun}},\
  }\bibfield  {title} {\bibinfo {title} {Spin phase regulated spin josephson
  supercurrent in topological superconductor},\ }\href
  {https://doi.org/10.1103/PhysRevB.105.184511} {\bibfield  {journal} {\bibinfo
   {journal} {Phys. Rev. B}\ }\textbf {\bibinfo {volume} {105}},\ \bibinfo
  {pages} {184511} (\bibinfo {year} {2022})}\BibitemShut {NoStop}%
\bibitem [{\citenamefont {Yang}\ \emph {et~al.}(2011)\citenamefont {Yang},
  \citenamefont {Xu}, \citenamefont {Sheng}, \citenamefont {Wang},
  \citenamefont {Xing},\ and\ \citenamefont {Sheng}}]{Yang2011}%
  \BibitemOpen
  \bibfield  {author} {\bibinfo {author} {\bibfnamefont {Y.}~\bibnamefont
  {Yang}}, \bibinfo {author} {\bibfnamefont {Z.}~\bibnamefont {Xu}}, \bibinfo
  {author} {\bibfnamefont {L.}~\bibnamefont {Sheng}}, \bibinfo {author}
  {\bibfnamefont {B.}~\bibnamefont {Wang}}, \bibinfo {author} {\bibfnamefont
  {D.~Y.}\ \bibnamefont {Xing}},\ and\ \bibinfo {author} {\bibfnamefont
  {D.~N.}\ \bibnamefont {Sheng}},\ }\bibfield  {title} {\bibinfo {title}
  {Time-reversal-symmetry-broken quantum spin hall effect},\ }\href
  {https://doi.org/10.1103/PhysRevLett.107.066602} {\bibfield  {journal}
  {\bibinfo  {journal} {Phys. Rev. Lett.}\ }\textbf {\bibinfo {volume} {107}},\
  \bibinfo {pages} {066602} (\bibinfo {year} {2011})}\BibitemShut {NoStop}%
\bibitem [{\citenamefont {Khalaf}\ \emph {et~al.}(2019)\citenamefont {Khalaf},
  \citenamefont {Kruchkov}, \citenamefont {Tarnopolsky},\ and\ \citenamefont
  {Vishwanath}}]{Khalaf2019}%
  \BibitemOpen
  \bibfield  {author} {\bibinfo {author} {\bibfnamefont {E.}~\bibnamefont
  {Khalaf}}, \bibinfo {author} {\bibfnamefont {A.~J.}\ \bibnamefont
  {Kruchkov}}, \bibinfo {author} {\bibfnamefont {G.}~\bibnamefont
  {Tarnopolsky}},\ and\ \bibinfo {author} {\bibfnamefont {A.}~\bibnamefont
  {Vishwanath}},\ }\bibfield  {title} {\bibinfo {title} {Magic angle hierarchy
  in twisted graphene multilayers},\ }\href
  {https://doi.org/10.1103/PhysRevB.100.085109} {\bibfield  {journal} {\bibinfo
   {journal} {Phys. Rev. B}\ }\textbf {\bibinfo {volume} {100}},\ \bibinfo
  {pages} {085109} (\bibinfo {year} {2019})}\BibitemShut {NoStop}%
\bibitem [{\citenamefont {Classen}\ \emph {et~al.}(2022)\citenamefont
  {Classen}, \citenamefont {Pixley},\ and\ \citenamefont
  {K{\"o}nig}}]{classen2022interaction}%
  \BibitemOpen
  \bibfield  {author} {\bibinfo {author} {\bibfnamefont {L.}~\bibnamefont
  {Classen}}, \bibinfo {author} {\bibfnamefont {J.}~\bibnamefont {Pixley}},\
  and\ \bibinfo {author} {\bibfnamefont {E.~J.}\ \bibnamefont {K{\"o}nig}},\
  }\bibfield  {title} {\bibinfo {title} {Interaction-induced velocity
  renormalization in magic-angle twisted multilayer graphene},\ }\href@noop {}
  {\bibfield  {journal} {\bibinfo  {journal} {2D Materials}\ }\textbf {\bibinfo
  {volume} {9}},\ \bibinfo {pages} {031001} (\bibinfo {year}
  {2022})}\BibitemShut {NoStop}%
\bibitem [{\citenamefont {Benalcazar}\ \emph {et~al.}(2017)\citenamefont
  {Benalcazar}, \citenamefont {Bernevig},\ and\ \citenamefont
  {Hughes}}]{benalcazar2017quantized}%
  \BibitemOpen
  \bibfield  {author} {\bibinfo {author} {\bibfnamefont {W.~A.}\ \bibnamefont
  {Benalcazar}}, \bibinfo {author} {\bibfnamefont {B.~A.}\ \bibnamefont
  {Bernevig}},\ and\ \bibinfo {author} {\bibfnamefont {T.~L.}\ \bibnamefont
  {Hughes}},\ }\bibfield  {title} {\bibinfo {title} {Quantized electric
  multipole insulators},\ }\href@noop {} {\bibfield  {journal} {\bibinfo
  {journal} {Science}\ }\textbf {\bibinfo {volume} {357}},\ \bibinfo {pages}
  {61} (\bibinfo {year} {2017})}\BibitemShut {NoStop}%
\bibitem [{\citenamefont {Schindler}\ \emph {et~al.}(2018)\citenamefont
  {Schindler}, \citenamefont {Cook}, \citenamefont {Vergniory}, \citenamefont
  {Wang}, \citenamefont {Parkin}, \citenamefont {Bernevig},\ and\ \citenamefont
  {Neupert}}]{schindler2018higher}%
  \BibitemOpen
  \bibfield  {author} {\bibinfo {author} {\bibfnamefont {F.}~\bibnamefont
  {Schindler}}, \bibinfo {author} {\bibfnamefont {A.~M.}\ \bibnamefont {Cook}},
  \bibinfo {author} {\bibfnamefont {M.~G.}\ \bibnamefont {Vergniory}}, \bibinfo
  {author} {\bibfnamefont {Z.}~\bibnamefont {Wang}}, \bibinfo {author}
  {\bibfnamefont {S.~S.}\ \bibnamefont {Parkin}}, \bibinfo {author}
  {\bibfnamefont {B.~A.}\ \bibnamefont {Bernevig}},\ and\ \bibinfo {author}
  {\bibfnamefont {T.}~\bibnamefont {Neupert}},\ }\bibfield  {title} {\bibinfo
  {title} {Higher-order topological insulators},\ }\href@noop {} {\bibfield
  {journal} {\bibinfo  {journal} {Science advances}\ }\textbf {\bibinfo
  {volume} {4}},\ \bibinfo {pages} {eaat0346} (\bibinfo {year}
  {2018})}\BibitemShut {NoStop}%
\end{thebibliography}%

\end{document}